\documentclass[ALICE,manyauthors]{cernphprep}

\usepackage[comma,square,numbers,sort&compress]{natbib}
\usepackage{hyperref}
\usepackage{color}
\usepackage{lineno}
\usepackage{booktabs}
\usepackage{xspace}

\DeclareGraphicsExtensions{.pdf} 
\graphicspath{{./figs/}}

\usepackage[T1]{fontenc}

\newcommand{\pt}{\ensuremath{p_{\mathrm T}}\xspace} 
\newcommand{\s}{\ensuremath{\sqrt{s}}}   
   
\newcommand{\sqrts}[1]   {\mbox{$\sqrt{s} = #1$ TeV}\xspace}

\newcommand{\mult}{charged-particle \mbox{multiplicity}\xspace}

\newcommand{\jpsi}{\mbox{J/$\psi$}\xspace}
\newcommand{\diele}{\ensuremath{{\mathrm e}^{+}{\mathrm e}^{-}}}

\newcommand{\dEdx}{\ensuremath{\text{d}E/\text{d}x}\xspace}

\newcommand{\dnjdy}{\ensuremath{{\mathrm d}N_{\text{J/$\psi$}}/{\mathrm d}y}\xspace}
\newcommand{\dndeta}{\ensuremath{{\mathrm d}N_{\mathrm{ch}}/{\mathrm d}\eta}\xspace}
\newcommand{\mdndeta}{\ensuremath{\langle {\mathrm d} N_{\mathrm{ch}}/{\mathrm d}\eta \rangle}\xspace}

\newcommand{\ntrcorr}{\ensuremath{N_{\mathrm{trk}}^{\mathrm{corr}}}\xspace}
\newcommand{\nch}{\ensuremath{N_{\mathrm{ch}}}\xspace}
\newcommand{\zvtx}{\ensuremath{z_{\mathrm{vtx}}}\xspace}

\newcommand{\mpt}{\ensuremath{\langle {p}_{\mathrm T} \rangle}\xspace}

\newcommand\gevc{GeV/$c$\xspace}
\newcommand\gevcc{GeV/$c^2$}
\newcommand{\average}[1]{\ensuremath{\langle #1 \rangle}\xspace}

\newcommand{ \be }{\begin{eqnarray}}
\newcommand{ \ee }{\end{eqnarray}}

\newcommand{ \mathrms }[1]{r.m.s.}

\begin{document}%

%%%%%%%%%%%%%%%  Title page %%%%%%%%%%%%%%%%%%%%%%%%
\begin{titlepage}
\PHyear{2020} %CERN-EP-2020-088
\PHnumber{088}      % required, will be obtained from PH
\PHdate{19 May}  % required, will be obtained from PH
%

%%% Put your own title + short title here:
%\title{\jpsi production as a function of charged-particle pseudorapidity density in pp collisions at \s~ = 13 TeV}
\title{Multiplicity dependence of inclusive \jpsi production at midrapidity in pp collisions at \s~ = 13 TeV}
\ShortTitle{Multiplicity dependence of inclusive \jpsi production at \s~ = 13 TeV}   % appears on right page headers

%%% Do not change the next lines
\Collaboration{ALICE Collaboration\thanks{See Appendix~\ref{app:collab} for the list of collaboration members}}
\ShortAuthor{ALICE Collaboration} % appears on left page headers, do not change

\begin{abstract}
Measurements of the inclusive \jpsi yield as a function of charged-particle pseudorapidity density \dndeta in pp collisions at \sqrts{13} with ALICE at the LHC are reported. 
The \jpsi meson yield is measured at midrapidity ($|y|<0.9$) in the dielectron channel, for events selected based on the charged-particle multiplicity at midrapidity ($|\eta|<1$) and at forward rapidity ($-3.7<\eta<-1.7$ and $2.8<\eta<5.1$); both observables are normalized to their corresponding averages in minimum bias events.
The increase of the normalized \jpsi yield with normalized \dndeta is significantly stronger than linear and dependent on the transverse momentum. The data are compared to theoretical predictions, which describe the observed trends well, albeit not always quantitatively.
\end{abstract}

\end{titlepage}
\setcounter{page}{2}

%
%
%%%%%%%%%%% put the body of the article here
\section{Introduction}
\label{sec:intro}
%!TEX root = ./JpsiMult_pp13.tex
Hadronic charmonium production at collider energies is a complex and not yet fully understood process, involving hard-scale processes, i.e.\@ the initial heavy-quark pair production, which can be described by means of perturbative quantum chromodynamics (pQCD), as well as soft-scale processes, i.e.\@ the subsequent binding into a color-neutral charmonium state. The latter stage is addressed via models which assume that it factorizes with respect to the perturbative early stage. The widely used non-relativistic QCD (NRQCD) effective theory \cite{Bodwin:1994jh} incorporates contributions from several hadronization mechanisms, like color-singlet or color-octet models (see Ref.~\cite{Lansberg:2019adr} for a recent review on models and Ref.\cite{Andronic:2015wma} for a comparison with data of Run 1 at the LHC).
The NRQCD formalism combined with a Color Glass Condensate (CGC) description of the incoming protons~\cite{Ma:2014mri} is a recent example of a comprehensive treatment of the transverse momentum \pt and rapidity dependent production, in particular extended down to zero transverse momentum.
Measurements of inclusive \jpsi production, as reported in this publication, contain a non-prompt contribution from bottom-hadron decays and the production of bottom quarks can be calculated in QCD pertubatively.

The event-multiplicity dependent production of charmonium and open charm hadrons in pp and p--Pb collisions are observables having the potential to give new insights on processes at the parton level and on the interplay between the hard and soft mechanisms in particle production and is widely studied at the LHC.
ALICE has studied the multiplicity dependence in pp collisions at \sqrts{7} of inclusive \jpsi production at mid- and forward rapidity~\cite{Abelev:2012rz}, and prompt \jpsi (including feed down from $\psi(2S)$ and $\chi_c$), non-prompt \jpsi and D-meson production at midrapidity~\cite{Adam:2015ota}. The general observation is an increase of open and hidden charm production with charged-particle multiplicity measured at midrapidity. For the \jpsi production, multiplicities of about 4 times the mean value were reached. The results are consistent with an approximately linear increase of the normalized yield  as a function of the normalized multiplicity (both observables are normalized to their corresponding averages in minimum bias events). For the D-meson production, normalized event multiplicities of about 6 were reached; a stronger than linear increase of D-meson production was observed at the highest multiplicities.
Observations made by the CMS Collaboration for $\Upsilon(\mathrm{nS})$ production at midrapidity at $\s = 2.76$ TeV indicate a linear increase with the event activity, when measuring it at forward rapidity, and a stronger than linear increase with the event activity measured at midrapidity~\cite{Chatrchyan:2013nza}.
At RHIC, a measurement of \jpsi production as a function of multiplicity was recently performed by the STAR Collaboration~\cite{Adam:2018jmp} for $\s = 0.2$ TeV, showing similar trends as observed in the LHC data.
The \jpsi production as a function of charged-particle multiplicity was studied also in p--Pb collisions, exhibiting significant differences for different ranges of rapidity of the \jpsi meson~\cite{Adamova:2017uhu,Acharya:2020giw}.
A clear correlation with the event multiplicity (and event shape) was experimentally established for the inclusive charged-particle production~\cite{Acharya:2019mzb} as well as for identified particles, including multi-strange hyperons~\cite{Acharya:2018orn}.

Several theoretical models, described briefly in Section~\ref{sec:meas}, predict a correlation of the normalized \jpsi production with the normalized event multiplicity which is stronger than linear. These include a coherent particle production model~\cite{Kopeliovich:2013yfa}, the percolation model~\cite{Ferreiro:2012fb}, the EPOS3 event generator~\cite{Werner:2013tya}, a CGC-complemented NRQCD model~\cite{Ma:2018bax}, the PYTHIA 8.2 event generator~\cite{Sjostrand:2014zea,Weber:2018ddv}, and the 3-Pomeron CGC model~\cite{Siddikov:2019xvf}.
While for instance multiparton interactions (as implemented in PYTHIA) play an important role in charm(onium) production, it is important to notice that the predicted correlation is, in all the models to first order, the result of a (\nch-dependent) reduction of the charged-particle multiplicity. Well known is the color string reconnection mechanism implemented in PYTHIA, but initial-state effects as in CGC models lead, with very different physics, similarly to a reduction in particle multiplicity. 

In this Letter, the measurements of the inclusive \jpsi yield as a function of charged-particle pseudorapidity density in pp collisions at \sqrts{13} are presented. The  measurements are performed in the dielectron channel at midrapidity with the ALICE detector at the LHC. The \pt-integrated and differential results are presented for minimum bias events as well as for events triggered on high multiplicity, which extend the multiplicity range up to 7 times the average multiplicity, and on the electromagnetic calorimeter signals, which allow to access \pt values up to 15-40~\gevc.
Section~\ref{sec:data} outlines the experimental setup and the data sample; Section~\ref{sec:ana} describes the analysis, while Section~\ref{sec:meas} presents the results; a brief summary and outlook are given in Section~\ref{sec:conc}.

\section{Experiment and data sample}
\label{sec:data}
%!TEX root = ./JpsiMult_pp13.tex

The reconstruction of $\jpsi$ in the $\diele$ decay channel at midrapidity is performed using the ALICE central barrel detectors, described in detail in Refs.~\cite{Abelev:2008aa, Abelev:2014ffa}. The setup is located in a solenoidal magnet providing a field of 0.5 T oriented along the beam direction. 

For this analysis, a minimum bias (MB) trigger, a high multiplicity (HM) trigger, and two triggers based on the deposited energy in the combined Electromagnetic Calorimeter (EMCal)  and the Di-jet Calorimeter (DCal)~\cite{Abeysekara:2010ze, Cortese:1121574, Allen:1272952} are employed. Both the MB and HM triggers are provided by the V0 detector, that consists of two forward scintillator arrays~\cite{Abbas:2013taa} covering the pseudorapidity ranges $-3.7<\eta<-1.7$ and $2.8<\eta<5.1$. The MB trigger signal consists of a coincident signal in both arrays, while the HM trigger requires a signal amplitude in the V0 arrays above a threshold which corresponds to the 0.1\% highest multiplicity events.
The EMCal and DCal are located back-to-back in azimuth and form a two-arm electromagnetic calorimeter.
While the EMCal detector covers $|\eta|<0.7$ over an azimuthal angle of $ 80^\circ < \varphi < 187^\circ$, the DCal covers $0.22 <|\eta|< 0.7$ for $260^\circ < \varphi < 320^\circ$ and $|\eta|<0.7$ for $320^\circ < \varphi < 327^\circ$.
As a consequence of identical construction, both have identical granularity and intrinsic energy resolution. In this paper, EMCal and DCal will be referred to together as EMCal.
The EMCal trigger consists of the sum of energy in a sliding window of $4\times4$ towers above a given threshold (a tower is the smallest segmentation of  the EMCal). In this data set, the trigger requires the presence of a cluster with a minimum energy of 9 GeV (EG1) or 4 GeV (EG2) in coincidence with the MB trigger condition. 

Tracks are reconstructed in the pseudorapidity range $|\eta|<0.9$ using the Inner Tracking System (ITS)~\cite{Aamodt:2010aa}, which consists of six layers of silicon detectors around the beam pipe, and the Time Projection Chamber (TPC)~\cite{Alme:2010ke}, a large cylindrical gas detector providing tracking and particle identification via specific ionisation energy loss $\dEdx$.
The first two layers of the ITS (covering ${|\eta| <  2.0}$ and ${|\eta| <  1.4}$), the Silicon Pixel Detector (SPD), are used for the charged-particle multiplicity measurement at midrapidity by counting tracklets, reconstructed from pairs of hits in the two SPD layers pointing to the collision vertex.

The results presented in this Letter are obtained using data recorded by ALICE during the LHC Run~2 data taking period for pp collisions at \sqrts{13}. The number of selected events and the corresponding integrated luminosities~\cite{ALICE-PUBLIC-2016-002} are listed in Table~\ref{tab:lumi} for the different triggers used in this analysis.
For the  analyzed data set, the maximum interaction rate was 260 kHz, and the maximum pileup probability was about 5$\times 10^{-3}$.

 \begin{table}[h!]
 \centering
\caption{ Number of selected events and corresponding integrated luminosities for the different triggers used in this analysis. } \label{tab:lumi}
 \begin{tabular}{ccccc}
\toprule
& \multicolumn{2}{c}{MB and HM triggers}   & \multicolumn{2}{c}{EMCal triggers} \\
\midrule
& MB  & HM & EG1 & EG2 \\
 Number of events & $1.25 \times 10^{9}$          & $0.64 \times 10^{9} $        & $82.4 \times 10^{6}$             & $120 \times 10^{6}$ \\
 Integrated luminosity       &$21.6\pm1.1\,\mathrm{nb}^{-1}$&$5.4\pm0.1\,\mathrm{pb}^{-1}$&$7.2\pm0.1\,\mathrm{pb}^{-1}$&$0.82\pm0.02\,\mathrm{pb}^{-1}$\\
\bottomrule
\end{tabular}

 \end{table}

\section{Analysis}
\label{sec:ana}
%!TEX root = ./JpsiMult_pp13.tex

In this work the inclusive production of \jpsi mesons is studied as a function of the pseudorapidity density of charged particles at midrapidity, \dndeta.
The \jpsi yield in a given multiplicity interval and in a given rapidity ($y$) range \dnjdy is normalized to the \jpsi yield in the INEL$>$0 event class, \average{\dnjdy}. The INEL$>$0 event class contains all events with at least 1 charged particle in $|\eta|<1$. In this ratio, most of the systematic uncertainties related to tracking and particle identification cancel.

\subsection{Event selection}
All events selected in this analysis are required to have a reconstructed collision vertex within the longitudinal interval $|\zvtx|<10$~cm in order to ensure uniform detector performance and one SPD tracklet in $|\eta|<1$. Beam-gas events are rejected using timing cuts with the V0 detector.
Pileup events are rejected using a vertex finding algorithm based on SPD tracklets~\cite{Abelev:2014ffa}, allowing the removal of events with 2 vertices. Because of the relatively small in-bunch pileup probability and the further event selection performed in the analysis, the fraction of remaining pileup is negligible in the minimum bias events sample and at most 2\% in the high multiplicity triggered sample.

\begin{figure}[htbp]
  \centering
    \includegraphics[width=0.49\linewidth]{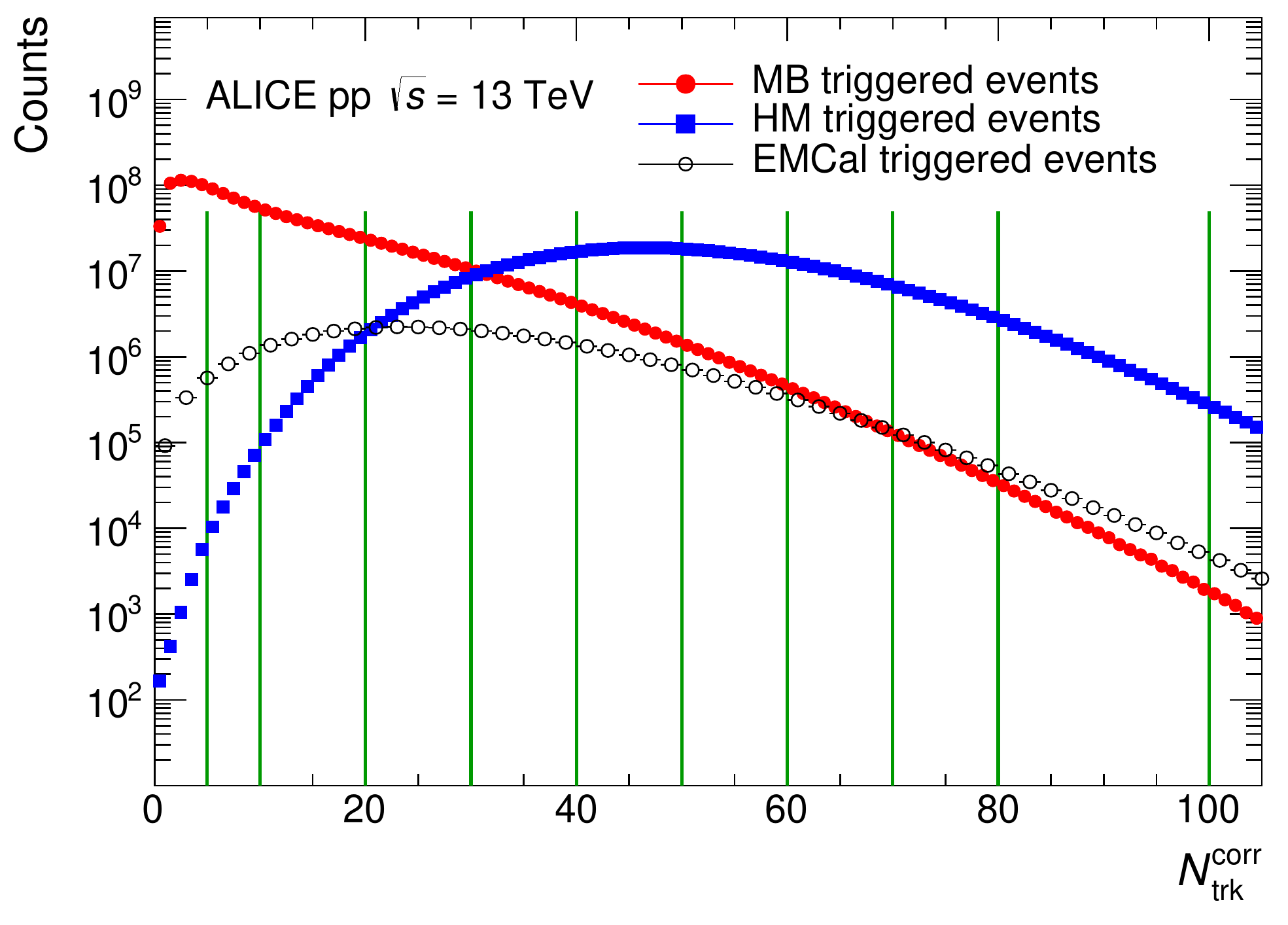}
    \includegraphics[width=0.49\linewidth]{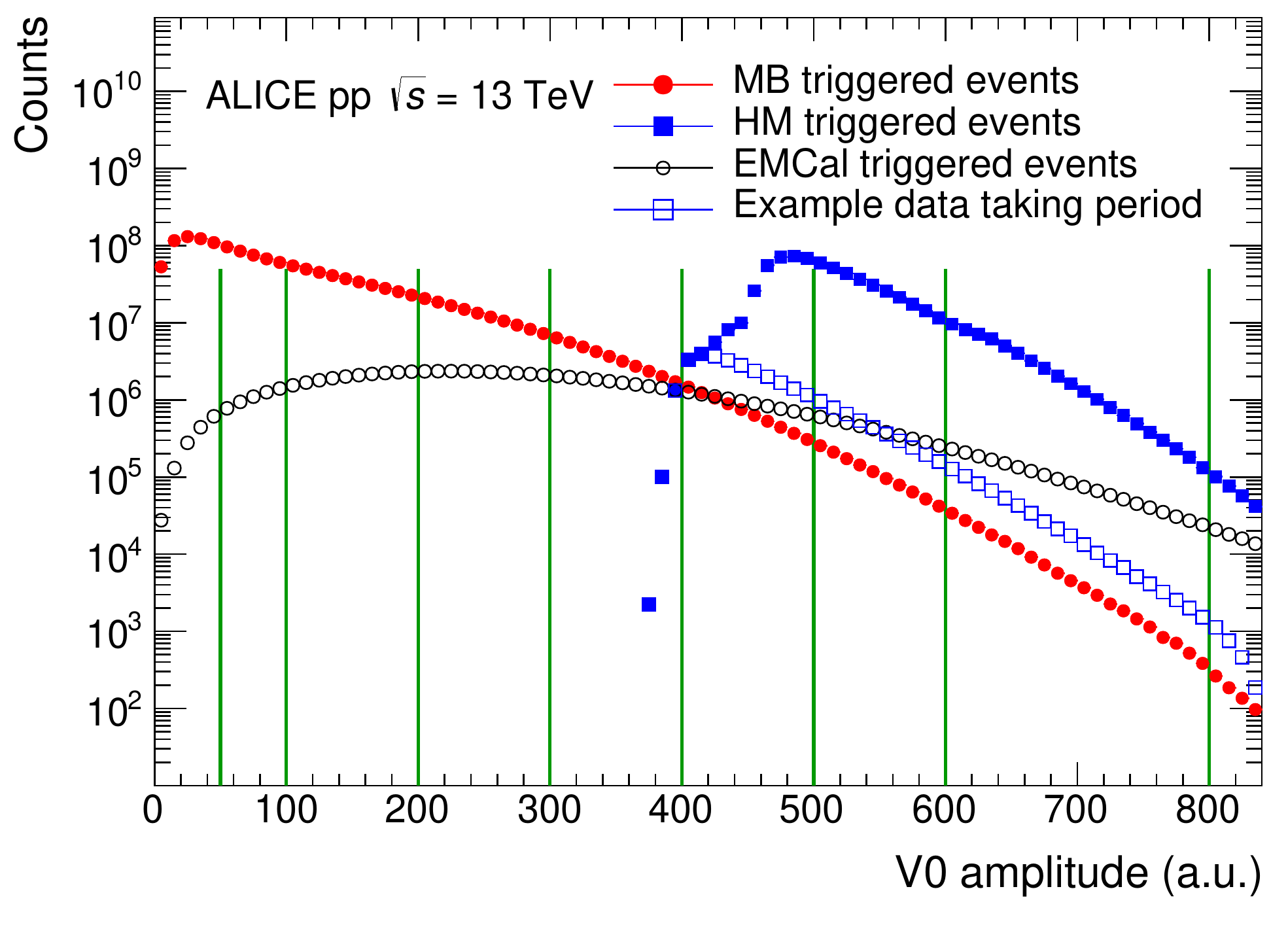}
  \caption{Distribution of the corrected SPD tracklets \ntrcorr (left) and V0 amplitude (right) for the MB events as well as the HM- and EMCal-triggered events used in the analysis. The vertical lines indicate the used multiplicity intervals (see Table~\ref{tab:mult}; the first bin spans from 0 to the position of the first line). For the HM-triggered events, the V0 amplitude distribution for a single data taking period is included for illustration (open squares).}
    \label{fig_nch_final}
\end{figure}

Events are binned in multiplicity classes based on either the SPD or the V0 detector signals, as shown in Fig.~\ref{fig_nch_final}. Events corresponding to the onset of the V0 HM trigger are excluded; %21% of HM triggers
that onset is rather sharp. The smearing seen in the distribution in the right panel of Fig.~\ref{fig_nch_final} is due to the different thresholds used during operation. To illustrate this, the V0-amplitude distribution for a single data taking period is included in Fig.~\ref{fig_nch_final} (right panel, open squares).

For the measurement of the charged-particle pseudorapidity density \dndeta at midrapidity, $|\eta|<1$, the SPD tracklets are used~\cite{Adam:2015pza}.
 Given the close proximity of the SPD detector to the interaction point (the two layers are at radial distances of 3.9 and 7.6 cm), its geometrical acceptance changes by up to ~50\% in the $\zvtx$ interval selected for analysis. In addition, the mean number of SPD tracklets also varied during the 3-year Run 2 data taking period due to changes in the number of active SPD detector elements. In order to compensate for these detector effects, a $\zvtx$ and time-dependent correction factor is applied such that the measured average multiplicity is equalized to a reference value. This reference was chosen to be the largest mean SPD tracklet multiplicity observed over time and $\zvtx$. This procedure is similar to what was done previously in Ref.~\cite{Abelev:2012rz}. The correction factor for each event is randomly smeared using a Poisson distribution to take into account event-by-event fluctuations. In the case of the event selection based on the forward multiplicity measurement with the V0 detector, the signal amplitudes are equalized to compensate for detector ageing and for the small acceptance variation with the event vertex position.

\begin{figure}[htbp]
  \centering
    \includegraphics[width=0.7\linewidth]{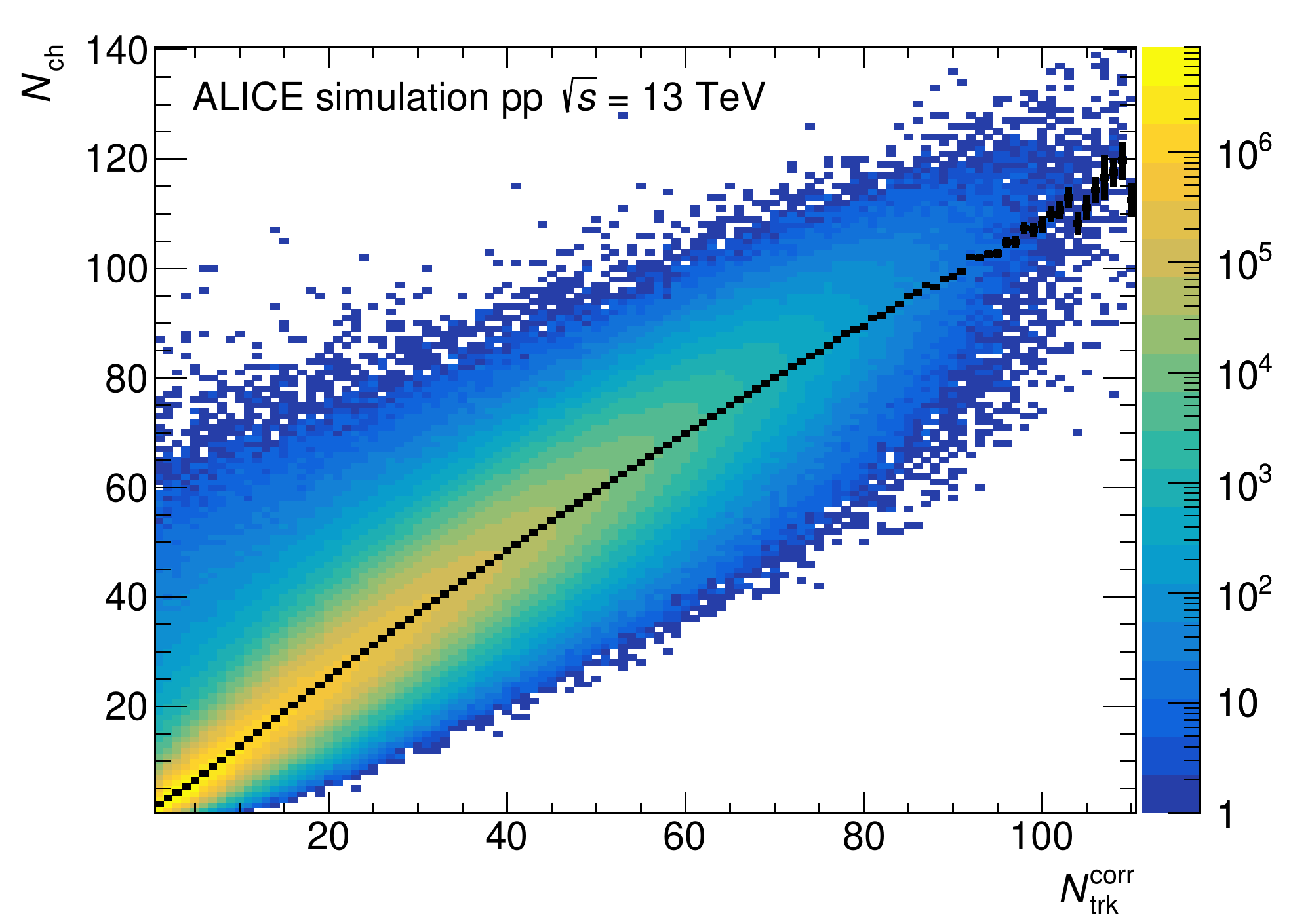}
  \caption{ Correlation between the number of generated primary charged particles, \nch, and the number of reconstructed SPD tracklets, \ntrcorr, in $|\eta|<1$, from PYTHIA 8.2 simulated collisions with detector transport through GEANT~3. The black points represent the mean values of \nch.}
    \label{fig_nch_ntrk}
\end{figure}

The overall inefficiency, the production of secondary particles due to interactions with the detector material and particle decays lead to a difference between the number of reconstructed tracklets and the true primary charged-particle multiplicity \nch ~(see details in Ref.~\cite{Adam:2015pza}).
Using  events simulated with the PYTHIA 8.2 event generator~\cite{Sjostrand:2007gs} (Monash 2013 tune, Ref.~\cite{Skands:2014pea}), the correlation between the tracklet multiplicity (after the $\zvtx$-correction), \ntrcorr, and the generated primary charged particles \nch ~is determined. The propagation of the simulated particles is done by GEANT~3~\cite{GEANT3} with a full simulation of the detector response, followed by the same reconstruction procedure as for real data.
The correction factor $\beta(\ntrcorr)=\nch/\ntrcorr$ to obtain the average \dndeta value corresponding to a given \ntrcorr bin is computed from the \ntrcorr--\nch ~correlation, shown in Fig.~\ref{fig_nch_ntrk} for events simulated with PYTHIA 8.2 and particle transport through GEANT~3.
As the generated charged-particle multiplicity in Monte Carlo differs from data, a corrected \nch distribution is constructed from the measured \ntrcorr distribution using Bayesian unfolding. From it, the corrected $\beta$ factors are obtained. A Monte Carlo closure test in PYTHIA 8.2 with unfolding based on EPOS-LHC events is used to validate the procedure.

The normalized charged-particle pseudorapidity density in each event class is calculated as:
\begin{equation}
\frac{\dndeta}{\average{\dndeta}_{\mathrm{INEL}>0}} = \frac{\beta \times \average{\ntrcorr}}{\Delta \eta \times \average{\dndeta}_{\mathrm{INEL}>0}},
\end{equation}
where \average{\ntrcorr} is the averaged value of \ntrcorr in each multiplicity class, corrected for the trigger and vertex finding efficiencies. The former is estimated from Monte Carlo simulations and the latter with a data driven approach. They are below unity only for the low-multiplicity events. The value corresponding to INEL$>$0 events, $\average{\dndeta}_{\mathrm{INEL}>0}$, was cross-checked with the published ALICE measurement~\cite{Adam:2015pza}, and is found to be in very good agreement. A similar procedure is also used for the event selection based on the V0 amplitude, measured as a sum of signals from charged particles in the intervals $-3.7<\eta<-1.7$ and $2.8<\eta<5.1$. The resulting values of the normalized multiplicity for the event classes considered in the analysis are summarised in Table~\ref{tab:mult} alongside the respective fractions of the INEL$>$0 cross section.

 \begin{table}[h!]
 \centering
 \caption{Average normalized charged-particle pseudorapidity density in $|\eta| <1$ for each event class selected in \ntrcorr measured in SPD ($|\eta| <1$; left part) and in V0 amplitude ($-3.7<\eta<-1.7$ and $2.8<\eta<5.1$; right part). The values correspond to the data sample used for the \pt-integrated analysis. Only systematic uncertainties are shown since the statistical ones are negligible. The corresponding fraction of the INEL$>$0 cross section for each event class is also indicated.} \label{tab:mult}
\begin{tabular}{cccc}
\toprule
\multicolumn{2}{c}{SPD selection} & \multicolumn{2}{c}{V0 selection} \\
\midrule
$\frac{\dndeta} {\average{\dndeta}_{\mathrm{INEL}>0}}$ & $\sigma / \sigma_{\textrm{INEL}>0}$ & $\frac{\dndeta }{ \average{\dndeta}_{\mathrm{INEL}>0}}$ & $\sigma / \sigma_{\textrm{INEL}>0}$\\
\midrule
 $ 0.23 \pm 0.01 $ & 32\% & $ 0.40 \pm 0.01 $ & 37\% \\
 $ 0.60 \pm 0.01 $ & 25\% & $ 0.76 \pm 0.01 $ & 26\% \\
 $ 1.23 \pm 0.02 $ & 25\% & $ 1.41 \pm 0.02 $ & 25\% \\
 $ 2.11 \pm 0.03 $ & 11\% & $ 2.26 \pm 0.03 $ & 9.0\% \\
 $ 2.98 \pm 0.05 $ & 4.7\% & $ 3.03 \pm 0.04 $ & 2.5\% \\
 $ 3.78 \pm 0.06 $ & 1.8\% & $ 3.92 \pm 0.06 $ & 0.5\% \\
 $ 4.58 \pm 0.08 $ & 0.6\% & $ 4.33 \pm 0.07 $ & 0.08\% \\
 $ 5.37 \pm 0.09 $ & 0.2\% & $ 4.96 \pm 0.08  $ & 0.01\% \\
 $ 6.17 \pm 0.11 $ & 0.05\% & ~ & ~ \\
 $ 7.13 \pm 0.12 $ & 0.02\% & ~ & ~ \\
\bottomrule
\end{tabular}
 \end{table}

\subsection{\jpsi signal extraction}
The \jpsi meson is measured in the dielectron decay channel at midrapidity. Electrons and positrons are reconstructed in the central barrel detectors by requiring a minimum of 70 out of maximally 159 track points in the TPC and a value of the track fit $\chi^2$ over the number of track points smaller than 4 \cite{Alme:2010ke}. Only tracks with at least two associated hits in the ITS, and one of them in the two innermost layers, are accepted. This requirement ensures both a good tracking resolution and the rejection of electrons and positrons produced from photons converting in the detector material. In the MB and HM trigger analysis, a further veto on the tracks belonging to identified photon conversion topologies is applied. The electron identification is achieved by the measurement of the specific energy loss of the track in the TPC, which is required to be compatible with that expected for electrons within 3 standard deviations. Tracks with a specific energy loss being consistent with that of the pion or proton hypothesis within 3.5 standard deviations are rejected. For the analysis of the EMCal-triggered events, the energy deposition of the track in the TPC is required to be in a range between $-$2.25 to $+$3 standard deviations around the mean expected value for the electrons. In addition, at least one of the \jpsi decay electrons is required to be matched to a cluster in the EMCal, with a cluster energy above the trigger threshold and an energy-to-momentum ratio in the range $0.8<E/p<1.3$.
Electrons and positrons are selected in the pseudorapidity range $|\eta|< 0.9$ and in the transverse momentum range $\pt > 1$ \gevc.

The number of reconstructed $\jpsi$ is obtained from the invariant mass distribution of all the opposite-sign (OS) pairs, which contains \diele pairs from \jpsi decays as well as combinatorics and other sources. In the MB and HM trigger analysis, the combinatorial background is estimated using a track rotation procedure in which one of the tracks is rotated by a random azimuthal angle multiple times to obtain a high statistics invariant mass distribution. This distribution is then normalized such that its integral over a range of the invariant mass well above the \jpsi~mass peak %(e.g. $[4.0,5.0]$ \gevcc)
matches the one of real OS pairs, and is subtracted from the latter distribution. The remaining residual background, which can be attributed to physical sources, e.g. correlated semileptonic decays of heavy-quark pairs, is estimated using a second-order polynomial function. 
For the analysis of the EMCal-triggered events, a fit to the OS invariant mass distribution is performed using a MC shape for the signal added to a polynomial to describe the background. A second- or third-order polynomial function is used, depending on the \pt range. 
The number of \jpsi is extracted by summing the dielectron yield in the background-subtracted invariant mass distribution in the mass interval $2.92<m_{\rm ee}<3.16$ \gevcc, which contains approximately 2/3 of the total reconstructed yield. The yield falling outside of the counting window at low invariant mass is due to the electron bremsstrahlung in the detector material and to the radiative \jpsi decay, and is corrected for using Monte Carlo simulations. Also, a correction for the yield loss due to the limited trigger and vertex finding efficiencies at low multiplicities is applied.

Due to the trigger enhancement, the yields obtained using the EMCal-triggered events were corrected by the trigger scaling factor, which is observed to be identical for all event classes. This correction is necessary to convert the yield per EMCal-triggered events into a yield per MB-triggered event and is accomplished by a data-driven method using the ratio of the cluster energy distribution in triggered data divided by the cluster energy distribution in minimum bias data. The ratio flattens above the trigger threshold and the scaling factor is then obtained by fitting a constant to the flat interval. 

\begin{figure}[th]
    \includegraphics[width=0.33\linewidth]{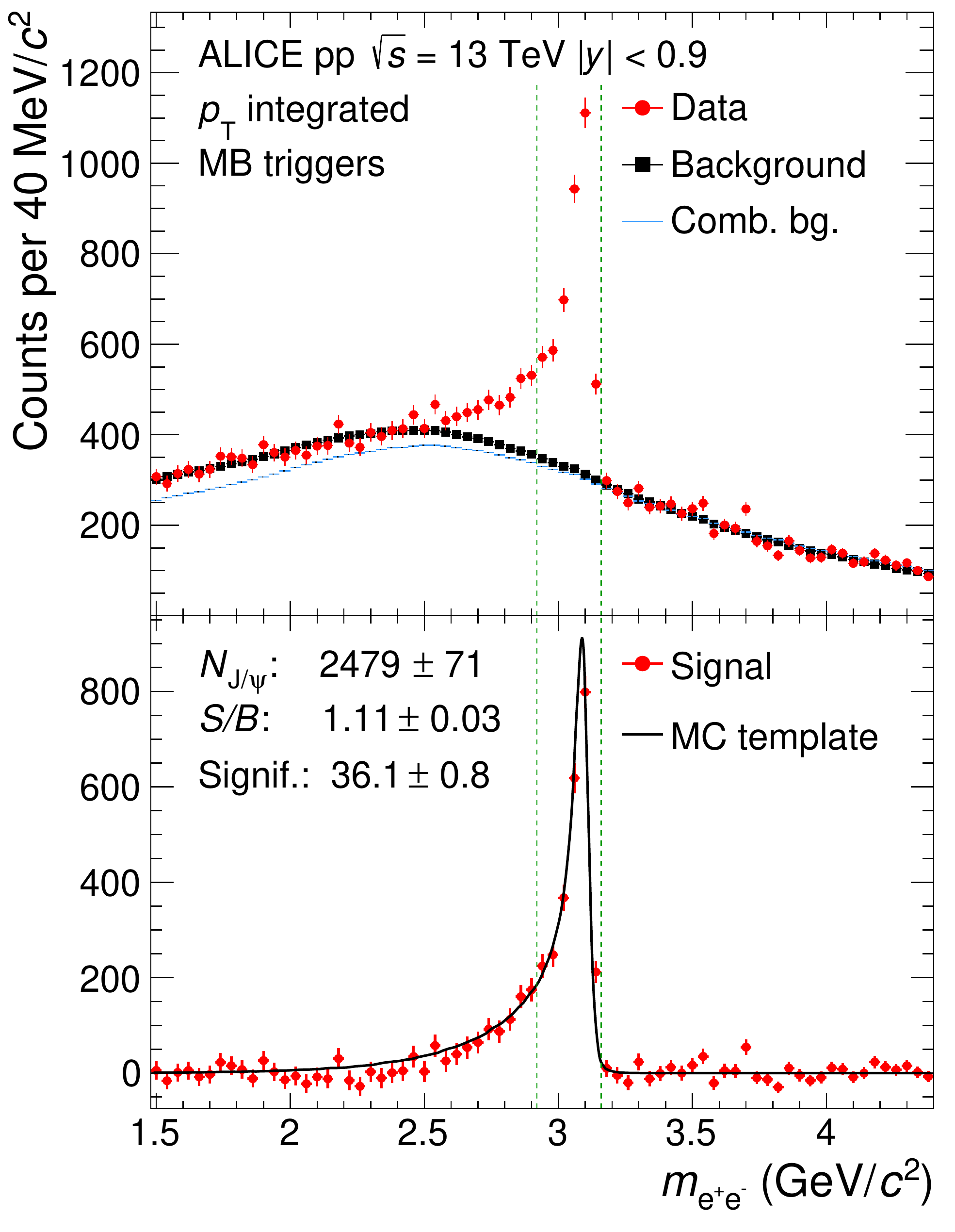}
    \includegraphics[width=0.33\linewidth]{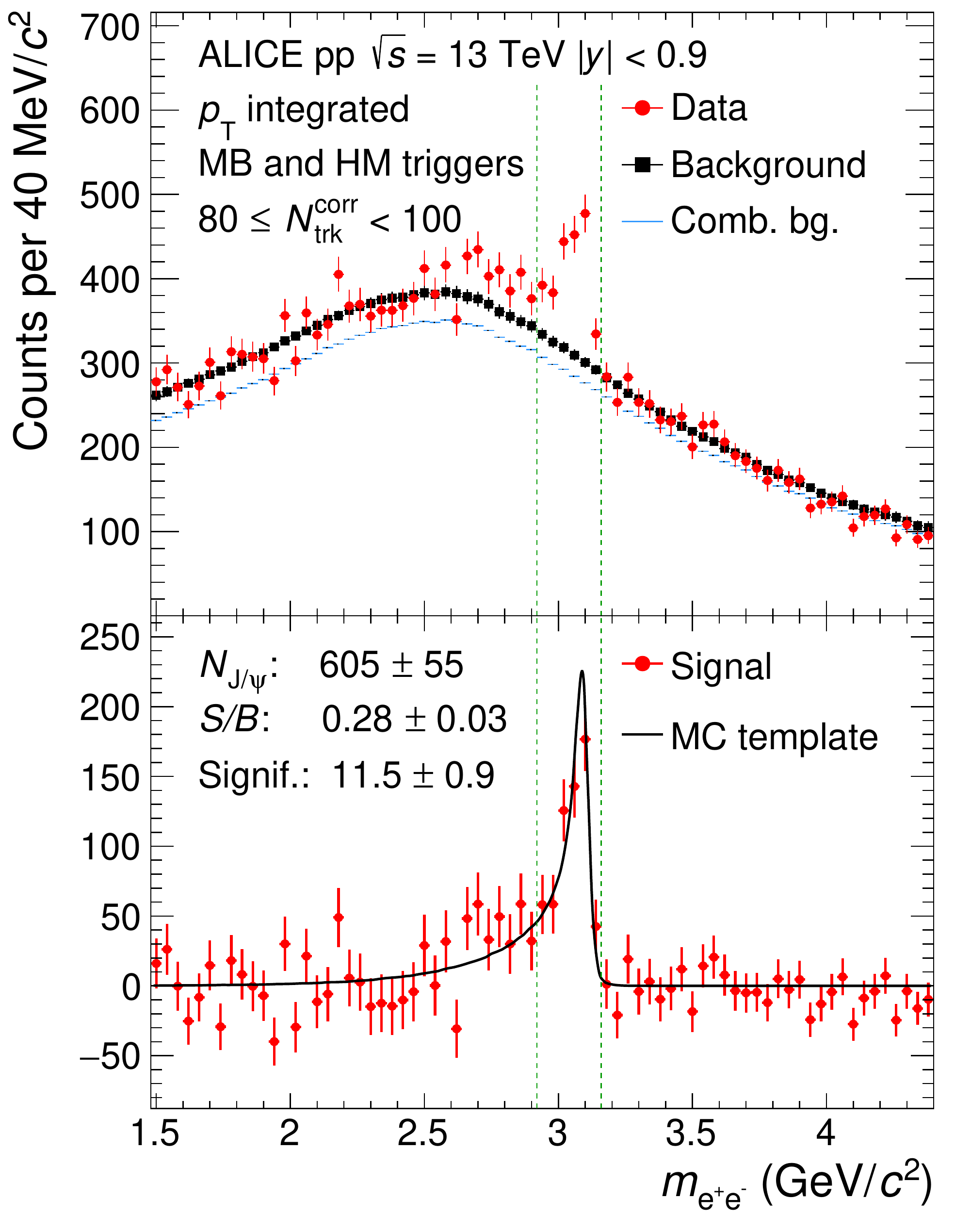}
    \includegraphics[width=0.33\linewidth]{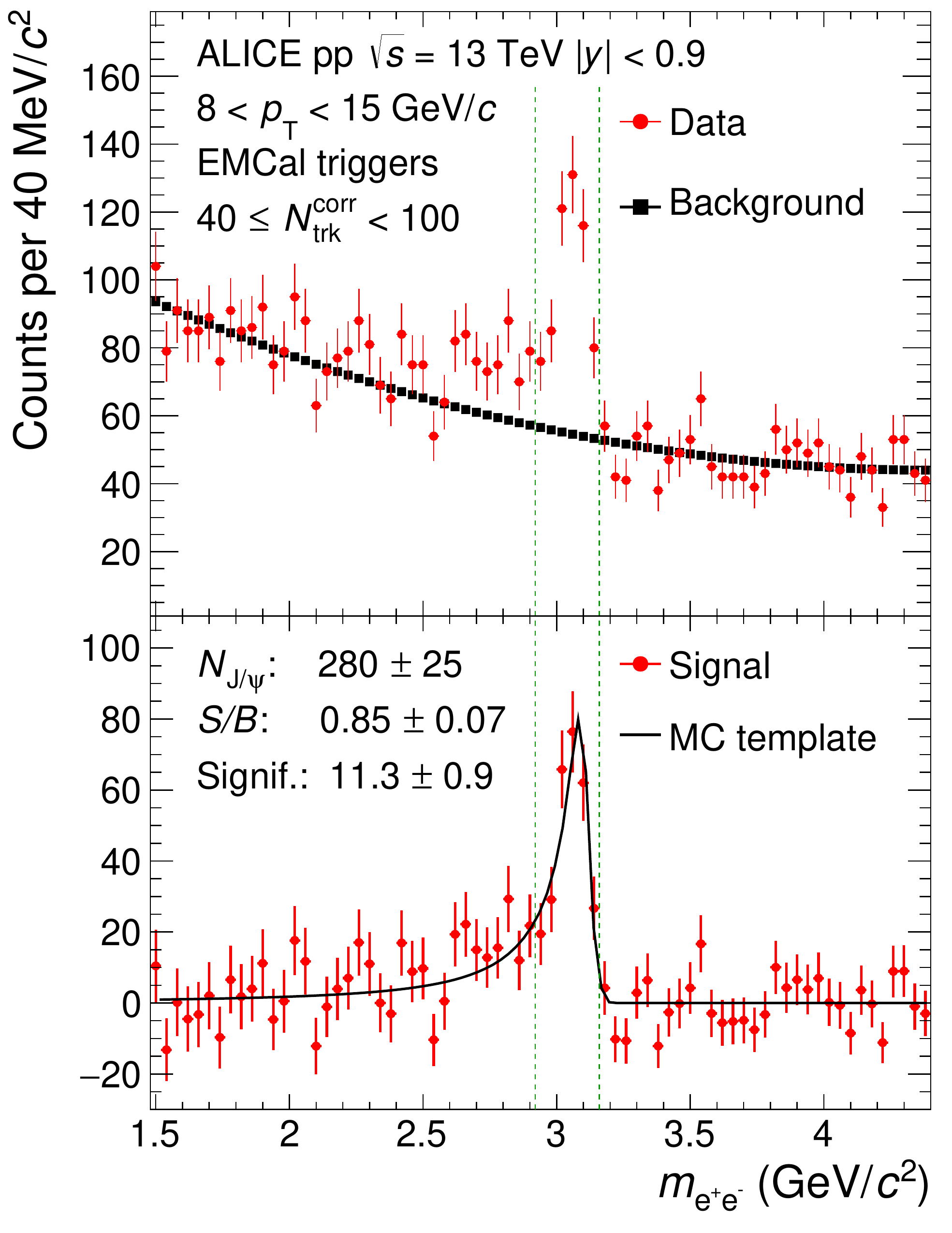}
  \caption{Top: Invariant mass distribution of electron-positron pairs for MB (left), HM (middle) and EMCal (right) triggers, together with combinatorial background estimation from the track-rotation method (blue lines in the left and middle panels) and the full background estimation (black squares). In the lower panels, the \jpsi signal obtained after background subtraction is shown together with the \jpsi signal shape from Monte Carlo simulations.
  The entries contain a correction for the relative efficiency (see text).
  The vertical lines indicate the mass range for the signal counting.}
    \label{fig_signal_integrated}
\end{figure}

In the top panels of Fig.~\ref{fig_signal_integrated} are shown the OS invariant mass distribution for MB events (left), a high multiplicity interval from the HM- (middle) and EMCal-triggered events (right), together with the estimated background distribution. The combinatorial background distribution from the track rotation method is shown in the left and middle panels with the blue lines, while the total background is shown as black squares in all the panels. The signal obtained after background subtraction is described well by the signal shape obtained from Monte Carlo simulations (discussed below); these MC templates have been scaled and overlaid on the data points in the bottom panels of Fig.~\ref{fig_signal_integrated}.

The \jpsi measurement is performed integrated in transverse momentum and in the \pt intervals 0 $< \pt <$ 4 \gevc and 4 $< \pt < $ 8 \gevc, using the MB and HM triggers. At higher \pt, the \jpsi mesons are reconstructed using the EMCal triggered events in the transverse momentum intervals 8 $< \pt <$ 15 \gevc and 15 $< \pt < $ 40 \gevc.
It was checked that the acceptance and efficiency for \jpsi reconstruction are not dependent on the event multiplicity. This was performed using pp collisions simulated with the PYTHIA 8.2 event generator with an injected \jpsi signal. The dielectron decay is simulated with the EvtGen package~\cite{Lange:2001uf} using PHOTOS~\cite{Barberio:1993qi} to describe the final-state radiation. The \jpsi mesons are assumed to be unpolarised consistent with measurements in pp collisions at the LHC~\cite{Acharya:2018uww}.

To account for the multiplicity dependence of the \pt spectrum of the \jpsi mesons, a correction for the relative efficiency, namely the efficiency for a given \pt value relative to the \pt-integrated value, is applied to each dielectron pair. This is contained in the invariant mass distributions shown in Fig.~\ref{fig_signal_integrated}.

\subsection{Systematic uncertainties}
\paragraph{Normalized multiplicity:}
The systematic uncertainty on the normalized multiplicity contains contributions from the trigger, vertex finding, and SPD efficiencies.
The first two contributions are estimated using alternative approaches: the trigger efficiency is calculated in a data-driven way, and for  the vertex finding efficiency Monte Carlo simulations are used. The differences to the values obtained with the default methods are taken as the systematic uncertainties. The contribution from the vertex finding efficiency is below 1\% (relative uncertainty) in all event classes, the one from the trigger efficiency reaches a maximum value of 1.3\% for the lowest multiplicity class.

In order to estimate uncertainties due to the SPD tracklet reconstruction efficiency, the number of corrected tracklets is scaled up and down by 3\%, which is the maximum observed discrepancy of the average number of SPD tracklets between data and Monte Carlo simulations. This uncertainty amounts to 3.6\% in the lowest multiplicity class, and to less than 1.5\% in all other classes. 
The uncertainty from the unfolding of the charged-particle multiplicity distribution is estimated by varying the number of iterations used in the Bayesian unfolding, as well as by using an alternative unfolding method~\cite{Malaescu:2009dm}. The uncertainty is found to be negligible.
All the aforementioned uncertainty sources are added in quadrature, leading to a total uncertainty on the normalized multiplicity of 3.7\% for the lowest multiplicity class, and to less than 2\% for all other classes.

\paragraph{Normalized \jpsi yield:}
The systematic uncertainties of the normalized \jpsi  yield are due to the signal extraction, bin-flow caused by the Poissonian smearing applied for the $\zvtx$-dependent correction of the SPD acceptance and vertex finding, trigger and SPD efficiencies.
For the analysis of the EMCal-triggered events, there is an additional component due to the matching of tracks to EMCal clusters and the electron identification via the $E/p$ measurement, which has a non-negligible multiplicity dependence. 
The  $E/p$ interval and the value of $E$ used to select only electrons above the EMCal trigger threshold are varied to determine the systematic uncertainty of the electron identification with the EMCal, leading to values from 1\% to 12\%, depending on the multiplicity bin.

The uncertainty of the \jpsi signal extraction is determined by varying the functions used to fit the background (first- or second-degree polynomials or exponential) and the fitting ranges, with the RMS of the distribution of normalized yields obtained from these variations being taken as a systematic uncertainty (the normalized yield corresponds to the default selection). The bin-flow effect is estimated from the RMS of the results obtained by repeating the analysis several times with different seeds for the random number generator.
The uncertainties from the signal extraction and the bin-flow effect are the dominant ones. They are of comparable size, with values between 1\% and 8\% depending on the multiplicity and \pt interval.
The uncertainties of the vertex finding, trigger and SPD tracklet efficiencies affect the estimated number of INEL$>$0 collisions, and hence the event-averaged minimum bias \jpsi yield \average{\dnjdy}, as well as the \jpsi yield in the low multiplicity classes. The uncertainties of the vertex finding and SPD efficiencies are below 1\% in most classes, while the uncertainty due to the trigger efficiency reaches up to 4\%, depending on the multiplicity class.

All the mentioned sources are added in quadrature to obtain the total systematic uncertainty, which, for the \pt-integrated results, varies between 3\% and 7\% with the multiplicity class. For the selected \pt intervals, the uncertainties are larger, varying between 3\% and 10\% with multiplicity and \pt interval, mainly due to the signal extraction, which is affected by statistical fluctuations of the background.
The results at high \pt, employing the EMCal, have uncertainties up to 13\%, which are larger because of the additional selection requirements on the track-cluster matching and the EMCal electron identification selections.

\section{Results and discussion}
\label{sec:meas}

%!TEX root = ./JpsiMult_pp13.tex

\begin{figure}[htb]
  \centering
    \includegraphics[width=0.49\linewidth]{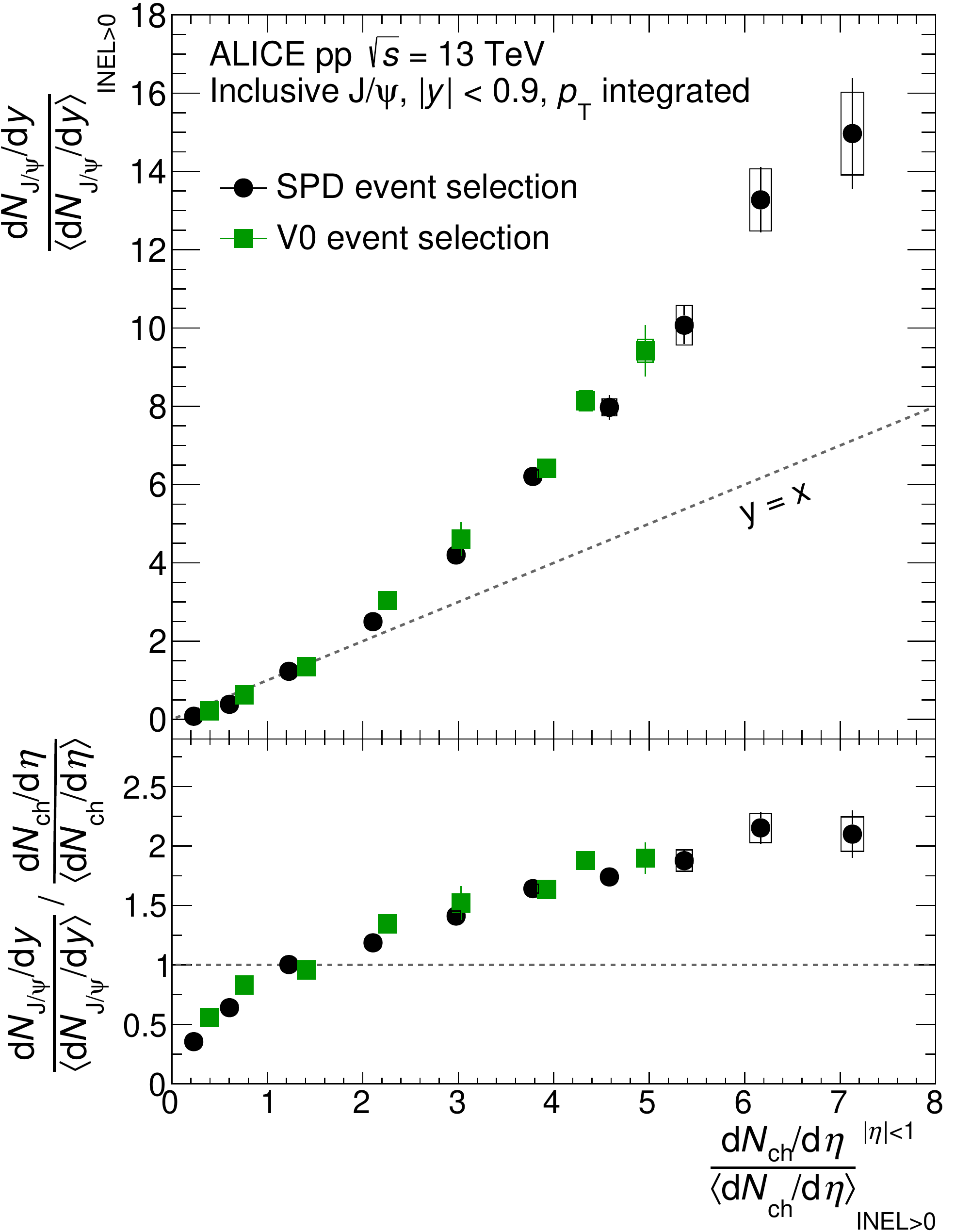}
  \caption{Normalized inclusive \pt-integrated \jpsi yield at midrapidity as a function of normalized charged-particle pseudorapidity density at midrapidity ($|\eta|<1$) with the event selection based on SPD tracklets at midrapidity and on V0 amplitude at forward rapidity in pp collisions at \sqrts{13}. Top: normalized \jpsi yield (diagonal drawn for reference). Bottom: double ratio of the normalized \jpsi yield and multiplicity. The error bars show statistical uncertainties and the boxes systematic uncertainties.}
    \label{fig_results_mid_fwd}
\end{figure}

The top panel of Fig. \ref{fig_results_mid_fwd} shows the normalized \jpsi yield as a function of the normalized charged-particle pseudorapidity density at midrapidity, $\dndeta / \average{\dndeta}$. The dashed line also shown in the figure is a linear function with the slope of unity. 

These results include both the MB and HM triggered events, which allow for a coverage of up to 7 times the average charged-particle multiplicity, 
when events are selected based on the measured midrapidity multiplicity. This is a significant extension with respect to our previous results in pp collisions at \sqrts{7}~\cite{Abelev:2012rz}, where only the range up to 4 was covered and with larger uncertainties. Using the event selection based on the V0 forward multiplicity (green squares), which should largely remove a possible auto-correlation bias, the measurement extends only up to 5 times the \mdndeta.
The results for the two event selection methods are in very good agreement. 
In both cases, the normalized \jpsi yield grows significantly faster than linear with the normalized multiplicity.

Included in Fig.~\ref{fig_results_mid_fwd} is also the double ratio of the normalized \jpsi yield to the normalized multiplicity (bottom panel). Two regimes could be identified, with a stronger increase of the double ratio for events with small multiplicity and a weaker increase for high-multiplicity events.
It is noted that the ``energy cost'' for the production of a \jpsi meson, characterized by a transverse mass $m_\mathrm T=\sqrt{m^2_{\mathrm J/\psi}+\pt^2/c^2}\simeq 5$ GeV/$c^2$, is similar to the one for particle production per unit rapidity of the underlying MB event, estimated as \mdndeta$\cdot$\mpt. A linear (diagonal) correlation with multiplicity is then expected to first order and observed in PYTHIA 8.2 simulations \cite{Weber:2018ddv}. As seen in Fig.~\ref{fig_results_mid_fwd}, the data exhibit richer features than this baseline expectation.

\begin{figure}[htb]
  \centering
    \includegraphics[width=0.45\linewidth]{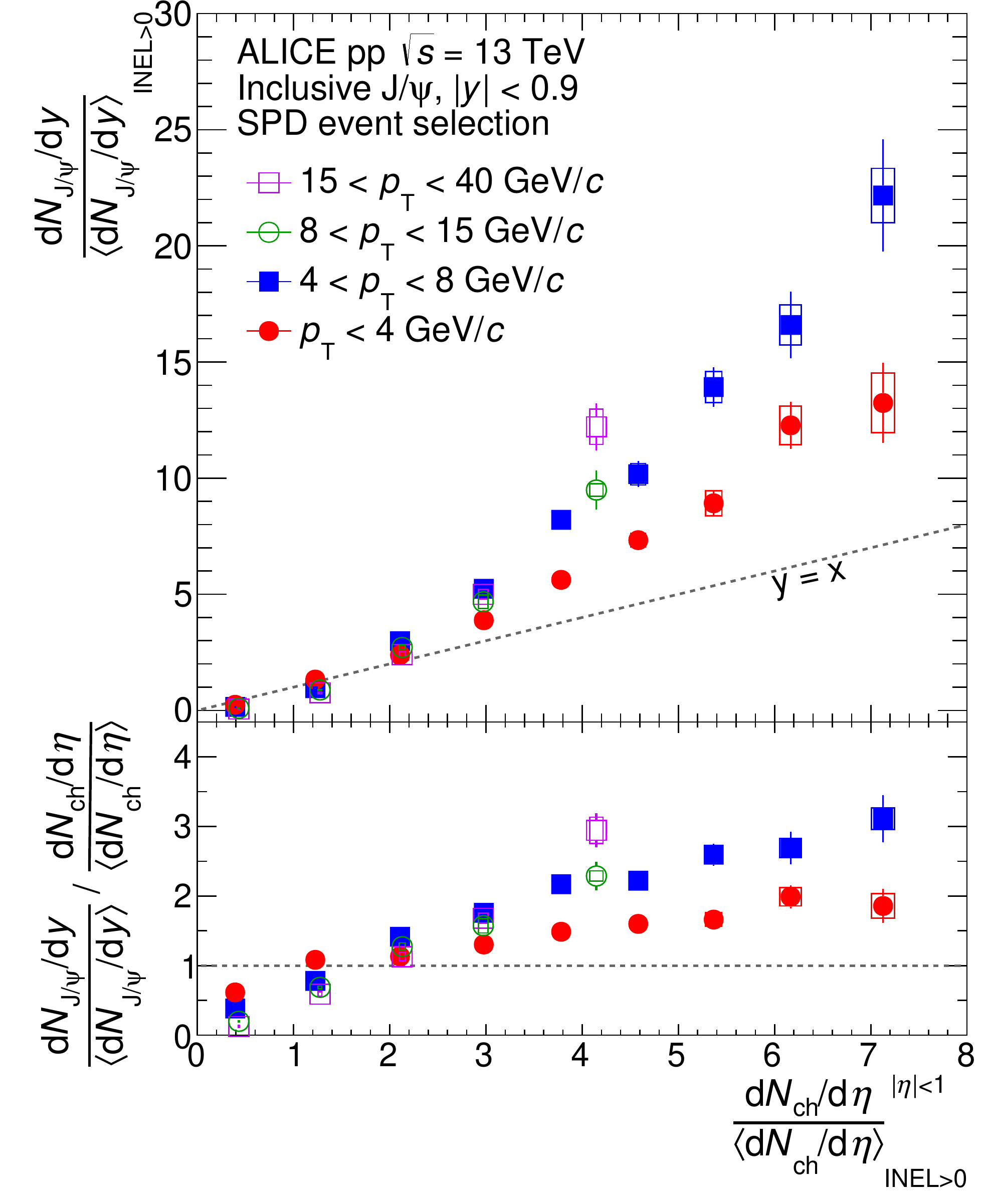}
    \includegraphics[width=0.45\linewidth]{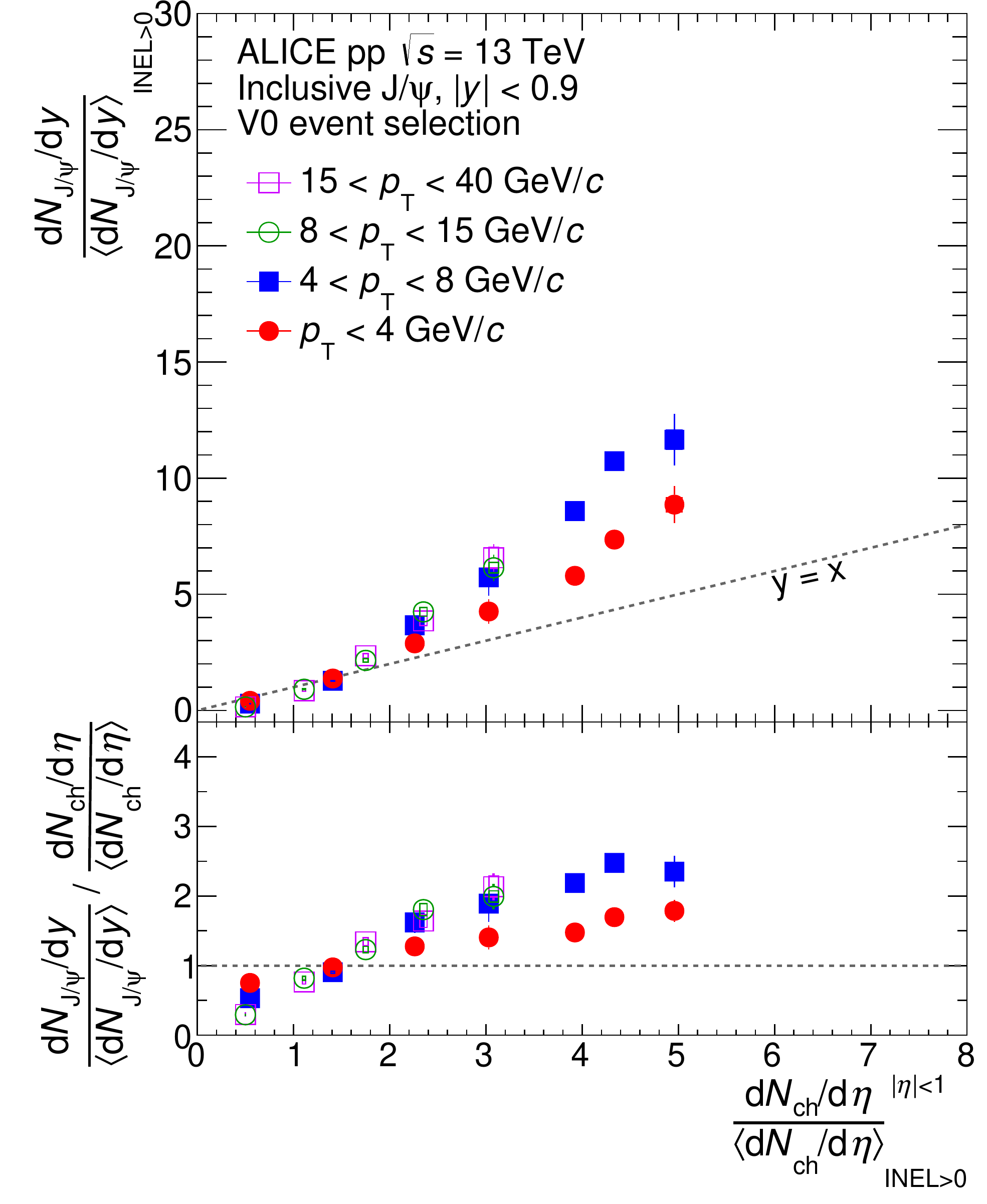}
  \caption{Normalized inclusive \jpsi yield at midrapidity as a function of normalized \mult in pp collisions at \sqrts{13}, for different ranges of \pt of the \jpsi meson. Left: event selection based on multiplicity at midrapidity. Right: event selection based on multiplicity at forward rapidity. The error bars show statistical uncertainties and the boxes systematic uncertainties.}
    \label{fig_results_emcal}
\end{figure}

The data in intervals of \pt of the \jpsi meson are shown in Fig.~\ref{fig_results_emcal}. The data exhibit a significant increase of the normalized \jpsi yield with the normalized multiplicity between the \jpsi \pt intervals 0--4 and 4--8 \gevc.
This effect could be attributed to various contributions~\cite{Weber:2018ddv}, like associated \jpsi production with other hadrons in jet fragmentation or from beauty-quark fragmentation, as the fraction of \jpsi from b-hadron decays increases with \pt~\cite{Abelev:2012gx}.

Measurements of the correlation with the event multiplicity for inclusive charged-particle production have identified similar trends~\cite{Acharya:2019mzb} as for the \jpsi \pt dependence.
The strength of this correlation is similar for \jpsi and for inclusive charged particles (dominated by pions) for \pt values giving a comparable $m_{\mathrm{T}}$ value. 
The production of strange hyperons at midrapidity was also observed to exhibit a correlation with event multiplicity in proportion to their mass~\cite{Acharya:2019kyh}; a strong correlation was also measured for the $\Upsilon$ mesons~\cite{Chatrchyan:2013nza}.

\begin{figure}[hbt]
  \centering                                     
    \includegraphics[width=0.47\linewidth]{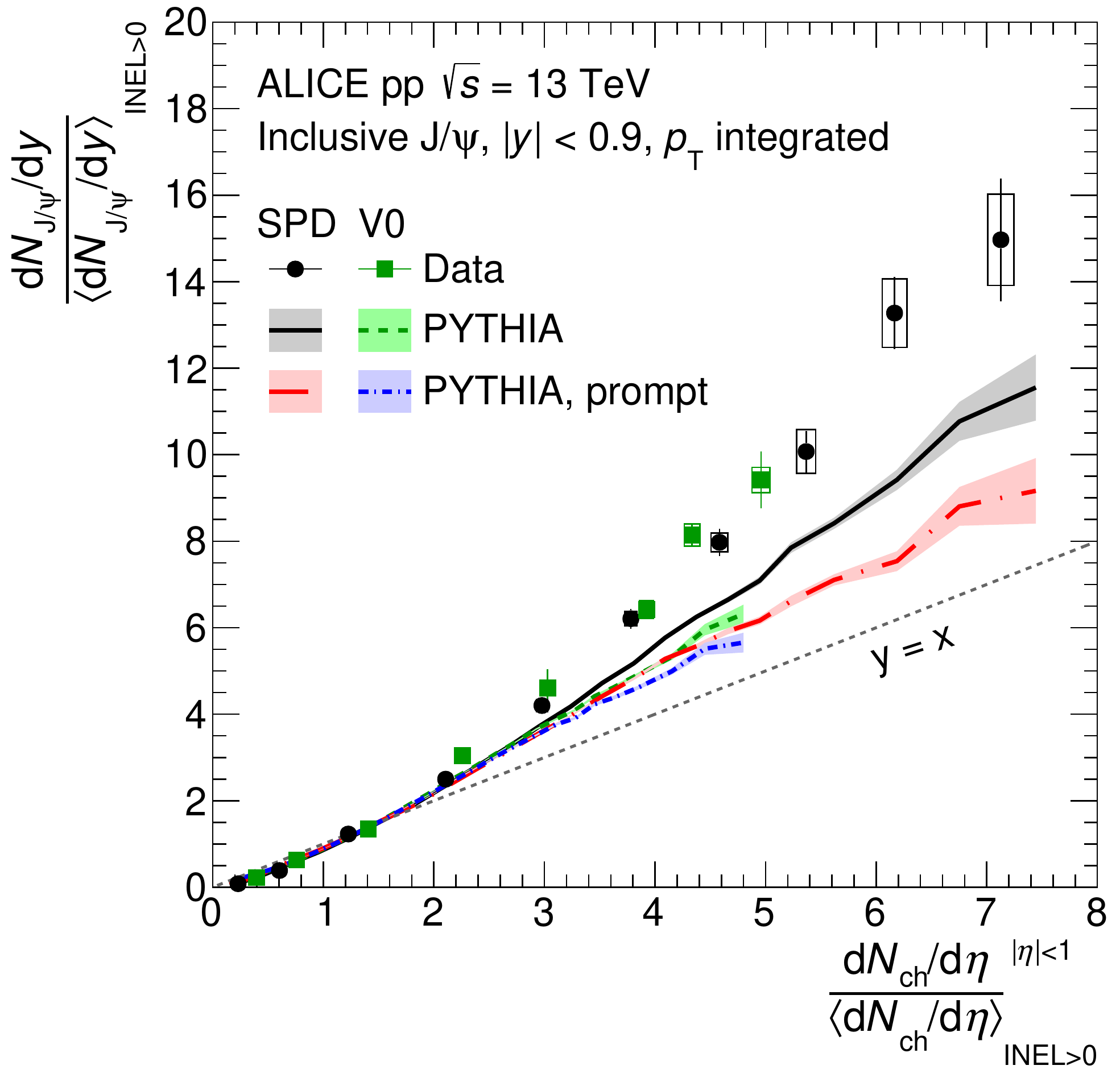}
    \includegraphics[width=0.47\linewidth]{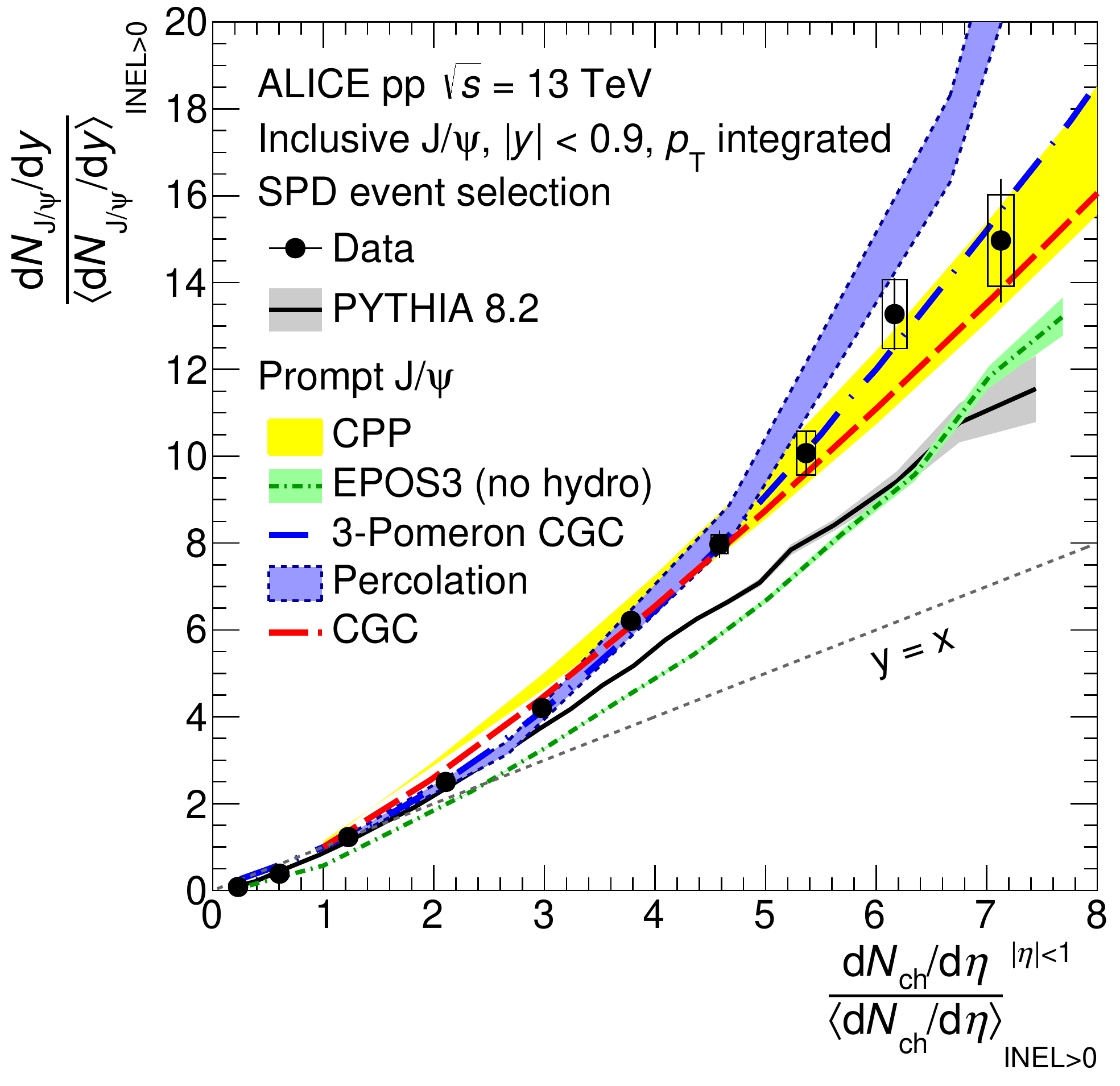}
  \caption{ Left: Comparison of data and PYTHIA 8.2 predictions for the two methods of event selection. For PYTHIA 8.2, the case of prompt J/$\psi$ meson production is included for illustration. Right: comparison of data (with SPD event selection) with model predictions from the coherent particle production model \cite{Kopeliovich:2013yfa}, the percolation model \cite{Ferreiro:2012fb}, the EPOS3 event generator \cite{Werner:2013tya}, the CGC model \cite{Ma:2018bax}, the 3-Pomeron CGC model \cite{Siddikov:2019xvf}, and PYTHIA 8.2 predictions. Except for the latter, none of the models include the non-prompt component.}
    \label{fig_results_models}
\end{figure}

The theoretical models currently available attribute the observed behavior to different underlying processes. In the PYTHIA 8.2 event generator~\cite{Sjostrand:2014zea}, multiparton interactions (MPI) are an important factor in charm-quark production. Indeed, from MPIs alone a stronger than linear scaling is expected for open-charm production, as was demonstrated in Ref.~\cite{Adam:2015ota} with PYTHIA 8.157. Taking into account all sources of heavy-quark production, however, a close to linear increase is predicted~\cite{Weber:2018ddv}. 
PYTHIA 8.2 reproduces well the observation in data with a very similar correlation with multiplicity for the two different rapidity intervals used for multiplicity measurement, as seen in the left panel of Fig.~\ref{fig_results_models}, although the overall slope of the trend is underestimated.
To illustrate the effect of non-prompt J/$\psi$ in the inclusive production, in Fig.~\ref{fig_results_models} the case of prompt J/$\psi$ meson production as predicted by PYTHIA 8.2 is shown in addition. A significant reduction of the correlation is observed, which appears to be stronger for the SPD event selection case.

In the EPOS3 event generator~\cite{Werner:2013tya,Werner:2010aa}, initial conditions are generated according to the parton-based Gribov-Regge formalism~\cite{Drescher:2000ha}. Sources of particle production in this framework are parton ladders, each composed of a pQCD hard process with initial- and final-state radiation. This model already predicted the stronger than linear increase with multiplicity observed for open-charm mesons~\cite{Adam:2015ota}, originating from  a collective (hydrodynamical) evolution of the system. The predictions from EPOS3, here without the hydrodynamic component, are similar in magnitude to those from PYTHIA 8.
In the percolation model~\cite{Ferreiro:2012fb}, spatially extended color strings are the sources of particle production in high-energy hadronic collisions. In a high-density environment they overlap; such a decrease in the effective number of strings leads to a reduction in particle production. Since the transverse size of a string is determined by its transverse mass, lighter particles are affected in a stronger way than heavier ones. This results in a linear increase of heavy-particle production at low multiplicities, gradually changing to a quadratic one at high multiplicities.
The coherent particle production (CPP) model \cite{Kopeliovich:2019phc,Kopeliovich:2013yfa} employs phenomenological parametrizations for light hadrons and J/$\psi$ derived from p--Pb collisions, and predicts a stronger than linear relative increase of \jpsi production with the normalized event multiplicity.
In the Color Glass Condensate (CGC) approach, the NRQCD framework is employed for \jpsi production. This effective field theory predicts, both for \jpsi and D mesons, a relative increase with the normalized multiplicity that is faster than linear, both for pp and p--Pb collisions~\cite{Ma:2018bax}. In a CGC saturation model, a faster than linear trend generically arises from the Bjorken-$x$ dependent saturation scale which would suppress more the soft-particle multiplicity, produced at low-$x$, compared to \jpsi production which is sensitive to larger values of $x$.
In the 3-Pomeron fusion model \cite{Siddikov:2019xvf}, the correlation arises as J/$\psi$ production via 3-gluon fusion processes from various Pomeron configurations are considered. The larger configuration space for the particular case of the overlapping rapidity interval for J/$\psi$ and charged particles leads to a significantly stronger correlation. Gluon saturation is implemented in this model; its effect, interestingly a reduced correlation, becomes significant for normalized multiplicities above 7.

All models predict an increase which is faster than linear, as shown in the right panel of Fig.\ref{fig_results_models}. In all models this is effectively the result of a (\nch-dependent) reduction of the charged-particle multiplicity, realized through rather different physics mechanisms in the various approaches (color string reconnection or percolation, gluon saturation, coherent particle production, 3-gluon fusion in gluon ladders/Pomerons).
The PYTHIA 8.2 and EPOS3 models underpredict the data, while the percolation model slightly overpredicts them at high multiplicity; good agreement is seen for the CGC, the coherent particle production, and the 3-Pomeron models.

\begin{figure}[htb]
  \centering
    \includegraphics[width=0.47\linewidth]{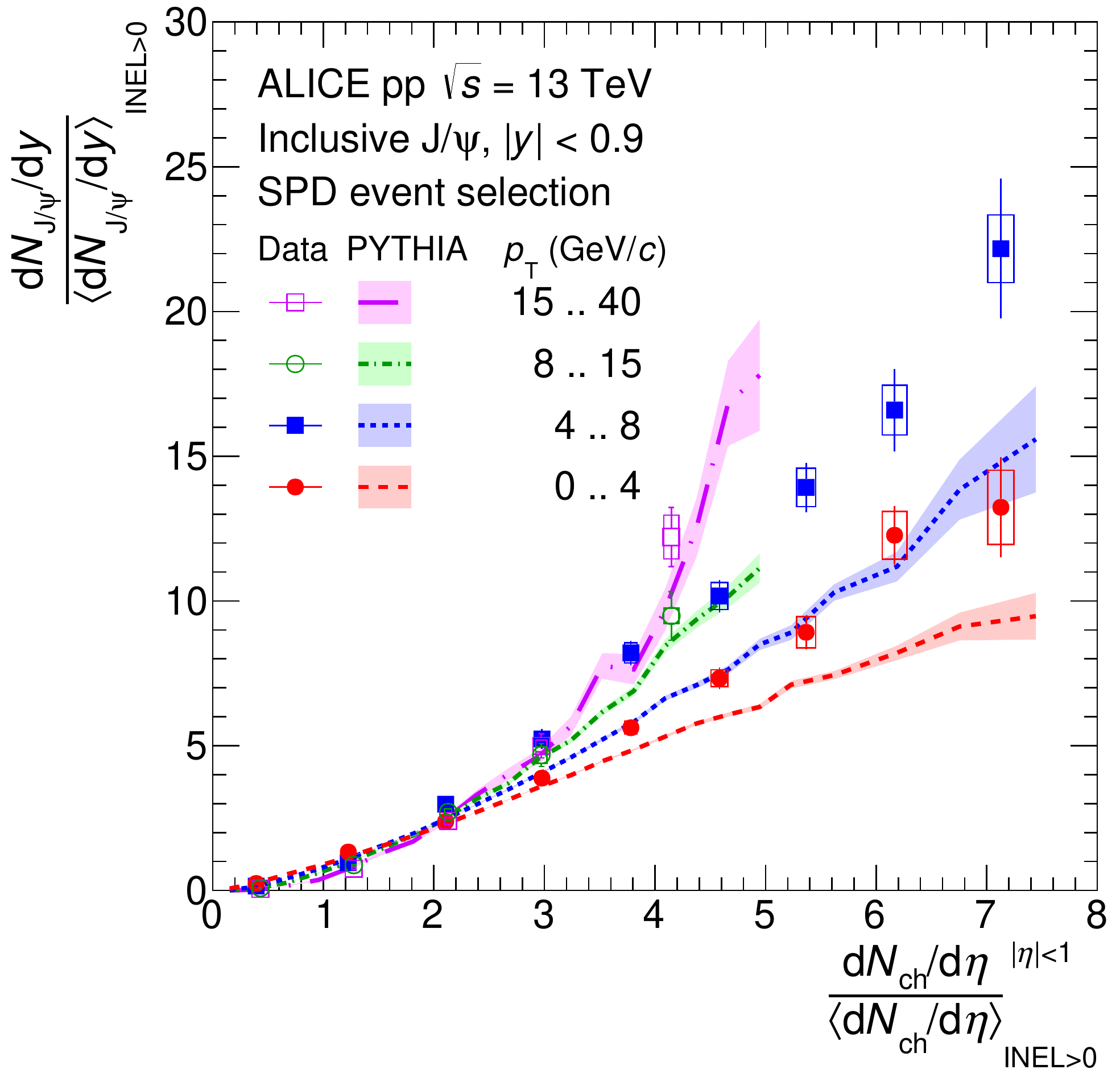}
	\caption{Normalized inclusive \jpsi yield at midrapidity as a function of normalized charged-particle pseudorapidity density at midrapidity for different \pt intervals; the data are compared to theoretical model predictions from PYTHIA 8.2. }
    \label{fig_results_models_pt}
\end{figure}

These observations need to be considered having in mind that in all models except PYTHIA 8.2 only the prompt \jpsi production is included. As illustrated in Fig.~\ref{fig_results_models} for PYTHIA 8.2, the prompt J/$\psi$ meson production exhibits a weaker relative increase with multiplicity compared to the inclusive production.
The agreement with data will improve in case of EPOS3 and will degrade for all the other models, in a consistent comparison. That could be realized either once the data for the prompt component will become available or as soon as the non-prompt component will be added to the current model predictions.

The contribution from decays of beauty hadrons increases significantly with \pt \cite{Abelev:2012gx} and might also have a different dependency on multiplicity; the existing measurement of charm and beauty production~\cite{Adam:2015ota} is not precise enough to be conclusive, but a study in PYTHIA 8.2~\cite{Weber:2018ddv} showed that the feed-down from beauty hadrons influences the result.
The trend of stronger increase in the \pt intervals above 4 \gevc seen in the data is qualitatively reproduced by PYTHIA 8.2, which, however, underestimates the data for $\pt < 8$ \gevc, as shown in Fig.~\ref{fig_results_models_pt}.

\section{Summary and conclusions}
\label{sec:conc}
%!TEX root = ./JpsiMult_pp13.tex

We have presented a comprehensive measurement of inclusive production of \jpsi mesons as a function of the event multiplicity in pp collisions at \sqrts{13} performed with the ALICE apparatus. The \jpsi production at midrapidity is studied using a data sample including minimum bias, high event activity, and EMCal triggered events. 
The event selection is performed based on the charged-particle measurement at midrapidity and in the forward region.
The \jpsi yield in a given multiplicity interval normalized to the \jpsi yield  in INEL$>$0 collisions is presented as a function of the charged-particle multiplicity similarly normalized.
The advantage of such a representation is that most of the experimental systematic uncertainties cancel; also, some of the theoretical model uncertainties are mitigated for such normalized yields.

A stronger than linear increase of the relative production of \jpsi as a function of multiplicity is observed for \pt-integrated yields; this increase is stronger for high-\pt \jpsi mesons. The trends are qualitatively, and for some of the models quantitatively, reproduced by theoretical models, but a critical appraisal of the similarity or difference between the physics mechanisms at play in various models is yet to be performed. More stringent tests of the models are needed too. Disentangling the feed-down from beauty hadrons, not included in most of the current theoretical predictions, will be an important step towards shedding light on the mechanism of hadronization of charm (and beauty) quarks, in particular in the environment of a high density of color strings created in high-multiplicity pp collisions. Data which will be collected in Run 3 at the LHC will be a significant addition for such studies.

%
%

%%%%% acknowledgements
\newenvironment{acknowledgement}{\relax}{\relax}
\begin{acknowledgement}
  \section*{Acknowledgements}

  We are grateful to E. Ferreiro, B. Kopeliovich, E. Levin, M. Siddikov, R. Venugopalan, K. Watanabe, and K. Werner for sending us the predictions of and clarifications about their models.
  
% Version: 2020-05-07

The ALICE Collaboration would like to thank all its engineers and technicians for their invaluable contributions to the construction of the experiment and the CERN accelerator teams for the outstanding performance of the LHC complex.
The ALICE Collaboration gratefully acknowledges the resources and support provided by all Grid centres and the Worldwide LHC Computing Grid (WLCG) collaboration.
The ALICE Collaboration acknowledges the following funding agencies for their support in building and running the ALICE detector:
A. I. Alikhanyan National Science Laboratory (Yerevan Physics Institute) Foundation (ANSL), State Committee of Science and World Federation of Scientists (WFS), Armenia;
Austrian Academy of Sciences, Austrian Science Fund (FWF): [M 2467-N36] and Nationalstiftung f\"{u}r Forschung, Technologie und Entwicklung, Austria;
Ministry of Communications and High Technologies, National Nuclear Research Center, Azerbaijan;
Conselho Nacional de Desenvolvimento Cient\'{\i}fico e Tecnol\'{o}gico (CNPq), Financiadora de Estudos e Projetos (Finep), Funda\c{c}\~{a}o de Amparo \`{a} Pesquisa do Estado de S\~{a}o Paulo (FAPESP) and Universidade Federal do Rio Grande do Sul (UFRGS), Brazil;
Ministry of Education of China (MOEC) , Ministry of Science \& Technology of China (MSTC) and National Natural Science Foundation of China (NSFC), China;
Ministry of Science and Education and Croatian Science Foundation, Croatia;
Centro de Aplicaciones Tecnol\'{o}gicas y Desarrollo Nuclear (CEADEN), Cubaenerg\'{\i}a, Cuba;
Ministry of Education, Youth and Sports of the Czech Republic, Czech Republic;
The Danish Council for Independent Research | Natural Sciences, the VILLUM FONDEN and Danish National Research Foundation (DNRF), Denmark;
Helsinki Institute of Physics (HIP), Finland;
Commissariat \`{a} l'Energie Atomique (CEA) and Institut National de Physique Nucl\'{e}aire et de Physique des Particules (IN2P3) and Centre National de la Recherche Scientifique (CNRS), France;
Bundesministerium f\"{u}r Bildung und Forschung (BMBF) and GSI Helmholtzzentrum f\"{u}r Schwerionenforschung GmbH, Germany;
General Secretariat for Research and Technology, Ministry of Education, Research and Religions, Greece;
National Research, Development and Innovation Office, Hungary;
Department of Atomic Energy Government of India (DAE), Department of Science and Technology, Government of India (DST), University Grants Commission, Government of India (UGC) and Council of Scientific and Industrial Research (CSIR), India;
Indonesian Institute of Science, Indonesia;
Centro Fermi - Museo Storico della Fisica e Centro Studi e Ricerche Enrico Fermi and Istituto Nazionale di Fisica Nucleare (INFN), Italy;
Institute for Innovative Science and Technology , Nagasaki Institute of Applied Science (IIST), Japanese Ministry of Education, Culture, Sports, Science and Technology (MEXT) and Japan Society for the Promotion of Science (JSPS) KAKENHI, Japan;
Consejo Nacional de Ciencia (CONACYT) y Tecnolog\'{i}a, through Fondo de Cooperaci\'{o}n Internacional en Ciencia y Tecnolog\'{i}a (FONCICYT) and Direcci\'{o}n General de Asuntos del Personal Academico (DGAPA), Mexico;
Nederlandse Organisatie voor Wetenschappelijk Onderzoek (NWO), Netherlands;
The Research Council of Norway, Norway;
Commission on Science and Technology for Sustainable Development in the South (COMSATS), Pakistan;
Pontificia Universidad Cat\'{o}lica del Per\'{u}, Peru;
Ministry of Science and Higher Education, National Science Centre and WUT ID-UB, Poland;
Korea Institute of Science and Technology Information and National Research Foundation of Korea (NRF), Republic of Korea;
Ministry of Education and Scientific Research, Institute of Atomic Physics and Ministry of Research and Innovation and Institute of Atomic Physics, Romania;
Joint Institute for Nuclear Research (JINR), Ministry of Education and Science of the Russian Federation, National Research Centre Kurchatov Institute, Russian Science Foundation and Russian Foundation for Basic Research, Russia;
Ministry of Education, Science, Research and Sport of the Slovak Republic, Slovakia;
National Research Foundation of South Africa, South Africa;
Swedish Research Council (VR) and Knut \& Alice Wallenberg Foundation (KAW), Sweden;
European Organization for Nuclear Research, Switzerland;
Suranaree University of Technology (SUT), National Science and Technology Development Agency (NSDTA) and Office of the Higher Education Commission under NRU project of Thailand, Thailand;
Turkish Atomic Energy Agency (TAEK), Turkey;
National Academy of  Sciences of Ukraine, Ukraine;
Science and Technology Facilities Council (STFC), United Kingdom;
National Science Foundation of the United States of America (NSF) and United States Department of Energy, Office of Nuclear Physics (DOE NP), United States of America.    %%%%%%% done by webmaster team
\end{acknowledgement}

%%%%%%%% Bibliography (In case of using bibtex generate the bbl requested by arXiv)
\bibliographystyle{utphys}   % Remember we use title in the biblio
\bibliography{Jpsi}

\providecommand{\href}[2]{#2}\begingroup\raggedright\begin{thebibliography}{10}

\bibitem{Bodwin:1994jh}
G.~T. Bodwin, E.~Braaten, and G.~Lepage, ``{Rigorous QCD analysis of inclusive
  annihilation and production of heavy quarkonium},''
  \href{http://dx.doi.org/10.1103/PhysRevD.55.5853}{{\em Phys. Rev. D}
  {\bfseries 51} (1995) 1125--1171},
  \href{http://arxiv.org/abs/hep-ph/9407339}{{\ttfamily arXiv:hep-ph/9407339}}.
  [Erratum: Phys.Rev.D 55, 5853 (1997)].

\bibitem{Lansberg:2019adr}
J.-P. Lansberg, ``{New Observables in Inclusive Production of Quarkonia},''
\href{http://arxiv.org/abs/1903.09185}{{\ttfamily arXiv:1903.09185 [hep-ph]}}.
%%CITATION = ARXIV:1903.09185;%%.

\bibitem{Andronic:2015wma}
A.~Andronic {\em et~al.}, ``{Heavy-flavour and quarkonium production in the LHC
  era: from proton-proton to heavy-ion collisions},''
  \href{http://dx.doi.org/10.1140/epjc/s10052-015-3819-5}{{\em Eur. Phys. J.}
  {\bfseries C76} no.~3, (2016) 107},
\href{http://arxiv.org/abs/1506.03981}{{\ttfamily arXiv:1506.03981 [nucl-ex]}}.
%%CITATION = ARXIV:1506.03981;%%.

\bibitem{Ma:2014mri}
Y.-Q. Ma and R.~Venugopalan, ``{Comprehensive description of \jpsi production
  in proton-proton collisions at collider energies},''
  \href{http://dx.doi.org/10.1103/PhysRevLett.113.192301}{{\em Phys. Rev.
  Lett.} {\bfseries 113} no.~19, (2014) 192301},
\href{http://arxiv.org/abs/1408.4075}{{\ttfamily arXiv:1408.4075 [hep-ph]}}.
%%CITATION = ARXIV:1408.4075;%%.

\bibitem{Abelev:2012rz}
{\bfseries ALICE} Collaboration, B.~Abelev {\em et~al.}, ``{$J/\psi$ production
  as a function of charged particle multiplicity in pp collisions at $\sqrt{s}
  = 7$ TeV},'' \href{http://dx.doi.org/10.1016/j.physletb.2012.04.052}{{\em
  Phys. Lett.} {\bfseries B712} (2012) 165--175},
\href{http://arxiv.org/abs/1202.2816}{{\ttfamily arXiv:1202.2816 [hep-ex]}}.
%%CITATION = ARXIV:1202.2816;%%.

\bibitem{Adam:2015ota}
{\bfseries ALICE} Collaboration, J.~Adam {\em et~al.}, ``{Measurement of charm
  and beauty production at central rapidity versus charged-particle
  multiplicity in proton-proton collisions at $ \sqrt{s}=7 $ TeV},''
  \href{http://dx.doi.org/10.1007/JHEP09(2015)148}{{\em JHEP} {\bfseries 09}
  (2015) 148},
\href{http://arxiv.org/abs/1505.00664}{{\ttfamily arXiv:1505.00664 [nucl-ex]}}.
%%CITATION = ARXIV:1505.00664;%%.

\bibitem{Chatrchyan:2013nza}
{\bfseries CMS} Collaboration, S.~Chatrchyan {\em et~al.}, ``{Event activity
  dependence of Y(nS) production in $\sqrt{s_{NN}}$ = 5.02 TeV pPb and
  $\sqrt{s}$ = 2.76 TeV pp collisions},''
  \href{http://dx.doi.org/10.1007/JHEP04(2014)103}{{\em JHEP} {\bfseries 04}
  (2014) 103},
\href{http://arxiv.org/abs/1312.6300}{{\ttfamily arXiv:1312.6300 [nucl-ex]}}.
%%CITATION = ARXIV:1312.6300;%%.

\bibitem{Adam:2018jmp}
{\bfseries STAR} Collaboration, J.~Adam {\em et~al.}, ``{$J/\psi$ production
  cross section and its dependence on charged-particle multiplicity in $p + p$
  collisions at $\sqrt{s}$ = 200 GeV},''
  \href{http://dx.doi.org/10.1016/j.physletb.2018.09.029}{{\em Phys. Lett.}
  {\bfseries B786} (2018) 87--93},
\href{http://arxiv.org/abs/1805.03745}{{\ttfamily arXiv:1805.03745 [hep-ex]}}.
%%CITATION = ARXIV:1805.03745;%%.

\bibitem{Adamova:2017uhu}
{\bfseries ALICE} Collaboration, D.~Adamov{\'a} {\em et~al.}, ``{J/$\psi$
  production as a function of charged-particle pseudorapidity density in p-Pb
  collisions at $\sqrt{s_{\rm NN}} = 5.02$ TeV},''
  \href{http://dx.doi.org/10.1016/j.physletb.2017.11.008}{{\em Phys. Lett.}
  {\bfseries B776} (2018) 91--104},
\href{http://arxiv.org/abs/1704.00274}{{\ttfamily arXiv:1704.00274 [nucl-ex]}}.
%%CITATION = ARXIV:1704.00274;%%.

\bibitem{Acharya:2020giw}
{\bfseries ALICE} Collaboration, S.~Acharya {\em et~al.}, ``{J/$\psi$
  production as a function of charged-particle multiplicity in p-Pb collisions
  at $\sqrt{\textit{s}_{\rm NN}}~=~8.16$ TeV},''
  \href{http://arxiv.org/abs/2004.12673}{{\ttfamily arXiv:2004.12673
  [nucl-ex]}}.

\bibitem{Acharya:2019mzb}
{\bfseries ALICE} Collaboration, S.~Acharya {\em et~al.}, ``{Charged-particle
  production as a function of multiplicity and transverse spherocity in pp
  collisions at $\sqrt{s} =5.02$ and 13 TeV},''
  \href{http://dx.doi.org/10.1140/epjc/s10052-019-7350-y}{{\em Eur. Phys. J.}
  {\bfseries C79} no.~10, (2019) 857},
\href{http://arxiv.org/abs/1905.07208}{{\ttfamily arXiv:1905.07208 [nucl-ex]}}.
%%CITATION = ARXIV:1905.07208;%%.

\bibitem{Acharya:2018orn}
{\bfseries ALICE} Collaboration, S.~Acharya {\em et~al.}, ``{Multiplicity
  dependence of light-flavor hadron production in pp collisions at $\sqrt{s}$ =
  7 TeV},'' \href{http://dx.doi.org/10.1103/PhysRevC.99.024906}{{\em Phys.
  Rev.} {\bfseries C99} no.~2, (2019) 024906},
\href{http://arxiv.org/abs/1807.11321}{{\ttfamily arXiv:1807.11321 [nucl-ex]}}.
%%CITATION = ARXIV:1807.11321;%%.

\bibitem{Kopeliovich:2013yfa}
B.~Z. Kopeliovich, H.~J. Pirner, I.~K. Potashnikova, K.~Reygers, and
  I.~Schmidt, ``{\jpsi in high-multiplicity pp collisions: Lessons from pA
  collisions},'' \href{http://dx.doi.org/10.1103/PhysRevD.88.116002}{{\em Phys.
  Rev.} {\bfseries D88} no.~11, (2013) 116002},
\href{http://arxiv.org/abs/1308.3638}{{\ttfamily arXiv:1308.3638 [hep-ph]}}.
%%CITATION = ARXIV:1308.3638;%%.

\bibitem{Ferreiro:2012fb}
E.~G. Ferreiro and C.~Pajares, ``{High multiplicity $pp$ events and $J/\psi$
  production at LHC},''
  \href{http://dx.doi.org/10.1103/PhysRevC.86.034903}{{\em Phys. Rev.}
  {\bfseries C86} (2012) 034903},
\href{http://arxiv.org/abs/1203.5936}{{\ttfamily arXiv:1203.5936 [hep-ph]}}.
%%CITATION = ARXIV:1203.5936;%%.

\bibitem{Werner:2013tya}
K.~Werner, B.~Guiot, I.~Karpenko, and T.~Pierog, ``{Analysing radial flow
  features in p-Pb and p-p collisions at several TeV by studying identified
  particle production in EPOS3},''
  \href{http://dx.doi.org/10.1103/PhysRevC.89.064903}{{\em Phys. Rev.}
  {\bfseries C89} no.~6, (2014) 064903},
\href{http://arxiv.org/abs/1312.1233}{{\ttfamily arXiv:1312.1233 [nucl-th]}}.
%%CITATION = ARXIV:1312.1233;%%.

\bibitem{Ma:2018bax}
Y.-Q. Ma, P.~Tribedy, R.~Venugopalan, and K.~Watanabe, ``{Event engineering
  studies for heavy flavor production and hadronization in high multiplicity
  hadron-hadron and hadron-nucleus collisions},''
  \href{http://dx.doi.org/10.1103/PhysRevD.98.074025}{{\em Phys. Rev.}
  {\bfseries D98} no.~7, (2018) 074025},
\href{http://arxiv.org/abs/1803.11093}{{\ttfamily arXiv:1803.11093 [hep-ph]}}.
%%CITATION = ARXIV:1803.11093;%%.

\bibitem{Sjostrand:2014zea}
T.~Sj{\"o}strand, S.~Ask, J.~R. Christiansen, R.~Corke, N.~Desai, P.~Ilten,
  S.~Mrenna, S.~Prestel, C.~O. Rasmussen, and P.~Z. Skands, ``{An Introduction
  to PYTHIA 8.2},'' \href{http://dx.doi.org/10.1016/j.cpc.2015.01.024}{{\em
  Comput. Phys. Commun.} {\bfseries 191} (2015) 159--177},
\href{http://arxiv.org/abs/1410.3012}{{\ttfamily arXiv:1410.3012 [hep-ph]}}.
%%CITATION = ARXIV:1410.3012;%%.

\bibitem{Weber:2018ddv}
S.~G. Weber, A.~Dubla, A.~Andronic, and A.~Morsch, ``{Elucidating the
  multiplicity dependence of $\mathrm {J}/\psi $ production in proton-proton
  collisions with PYTHIA8},''
  \href{http://dx.doi.org/10.1140/epjc/s10052-018-6531-4}{{\em Eur. Phys. J.}
  {\bfseries C79} no.~1, (2019) 36},
\href{http://arxiv.org/abs/1811.07744}{{\ttfamily arXiv:1811.07744 [nucl-th]}}.
%%CITATION = ARXIV:1811.07744;%%.

\bibitem{Siddikov:2019xvf}
E.~Levin, I.~Schmidt, and M.~Siddikov, ``{Multiplicity dependence of quarkonia
  production in the CGC approach},''
  \href{http://dx.doi.org/10.1140/epjc/s10052-020-8086-4}{{\em Eur. Phys. J. C}
  {\bfseries 80} no.~6, (2020) 560},
  \href{http://arxiv.org/abs/1910.13579}{{\ttfamily arXiv:1910.13579
  [hep-ph]}}.

\bibitem{Abelev:2008aa}
{\bfseries ALICE} Collaboration, B.~Abelev {\em et~al.}, ``The {ALICE}
  experiment at the {CERN} {LHC},'' {\em JINST} {\bfseries 3} (2008) S08002.

\bibitem{Abelev:2014ffa}
{\bfseries ALICE} Collaboration, B.~Abelev {\em et~al.}, ``{Performance of the
  ALICE Experiment at the CERN LHC},''
  \href{http://dx.doi.org/10.1142/S0217751X14300440}{{\em Int. J. Mod. Phys.}
  {\bfseries A29} (2014) 1430044},
\href{http://arxiv.org/abs/1402.4476}{{\ttfamily arXiv:1402.4476 [nucl-ex]}}.
%%CITATION = ARXIV:1402.4476;%%.

\bibitem{Abeysekara:2010ze}
{\bfseries ALICE EMCal} Collaboration, U.~Abeysekara {\em et~al.}, ``{ALICE
  EMCal Physics Performance Report},''
  \href{http://arxiv.org/abs/1008.0413}{{\ttfamily arXiv:1008.0413
  [physics.ins-det]}}.

\bibitem{Cortese:1121574}
{\bfseries ALICE} Collaboration, P.~Cortese {\em et~al.}, ``{ALICE
  Electromagnetic Calorimeter Technical Design Report},'' Tech. Rep.
  CERN-LHCC-2008-014. ALICE-TDR-14, Aug, 2008.
\newblock \url{https://cds.cern.ch/record/1121574}.

\bibitem{Allen:1272952}
J.~Allen {\em et~al.}, ``{ALICE DCal: An Addendum to the EMCal Technical Design
  Report Di-Jet and Hadron-Jet correlation measurements in ALICE},'' Tech. Rep.
  CERN-LHCC-2010-011. ALICE-TDR-14-add-1, Jun, 2010.
\newblock \url{https://cds.cern.ch/record/1272952}.

\bibitem{Abbas:2013taa}
{\bfseries ALICE} Collaboration, E.~Abbas {\em et~al.}, ``{Performance of the
  ALICE VZERO system},''
  \href{http://dx.doi.org/10.1088/1748-0221/8/10/P10016}{{\em JINST} {\bfseries
  8} (2013) P10016},
\href{http://arxiv.org/abs/1306.3130}{{\ttfamily arXiv:1306.3130 [nucl-ex]}}.
%%CITATION = ARXIV:1306.3130;%%.

\bibitem{Aamodt:2010aa}
{\bfseries ALICE} Collaboration, K.~Aamodt {\em et~al.}, ``{Alignment of the
  ALICE Inner Tracking System with cosmic-ray tracks},''
  \href{http://dx.doi.org/10.1088/1748-0221/5/03/P03003}{{\em JINST} {\bfseries
  5} (2010) P03003},
\href{http://arxiv.org/abs/1001.0502}{{\ttfamily arXiv:1001.0502
  [physics.ins-det]}}.
%%CITATION = ARXIV:1001.0502;%%.

\bibitem{Alme:2010ke}
J.~Alme {\em et~al.}, ``{The ALICE TPC, a large 3-dimensional tracking device
  with fast readout for ultra-high multiplicity events},''
  \href{http://dx.doi.org/10.1016/j.nima.2010.04.042}{{\em Nucl. Instrum.
  Meth.} {\bfseries A622} (2010) 316--367},
\href{http://arxiv.org/abs/1001.1950}{{\ttfamily arXiv:1001.1950
  [physics.ins-det]}}.
%%CITATION = ARXIV:1001.1950;%%.

\bibitem{ALICE-PUBLIC-2016-002}
{\bfseries ALICE} Collaboration, J.~Adam {\em et~al.}, ``{ALICE luminosity
  determination for pp collisions at $\sqrt{s}=13$ TeV},'' {\em
  ALICE-PUBLIC-2016-002} (Jun, 2016) .
  \url{https://cds.cern.ch/record/2160174}.

\bibitem{Adam:2015pza}
{\bfseries ALICE} Collaboration, J.~Adam {\em et~al.}, ``{Pseudorapidity and
  transverse-momentum distributions of charged particles in proton-proton
  collisions at $\sqrt s=$ 13 TeV},''
  \href{http://dx.doi.org/10.1016/j.physletb.2015.12.030}{{\em Phys. Lett.}
  {\bfseries B753} (2016) 319--329},
\href{http://arxiv.org/abs/1509.08734}{{\ttfamily arXiv:1509.08734 [nucl-ex]}}.
%%CITATION = ARXIV:1509.08734;%%.

\bibitem{Sjostrand:2007gs}
T.~Sj{\"o}strand, S.~Mrenna, and P.~Z. Skands, ``{A Brief Introduction to
  PYTHIA 8.1},'' \href{http://dx.doi.org/10.1016/j.cpc.2008.01.036}{{\em
  Comput. Phys. Commun.} {\bfseries 178} (2008) 852--867},
\href{http://arxiv.org/abs/0710.3820}{{\ttfamily arXiv:0710.3820 [hep-ph]}}.
%%CITATION = ARXIV:0710.3820;%%.

\bibitem{Skands:2014pea}
P.~Skands, S.~Carrazza, and J.~Rojo, ``{Tuning PYTHIA 8.1: the Monash 2013
  Tune},'' \href{http://dx.doi.org/10.1140/epjc/s10052-014-3024-y}{{\em Eur.
  Phys. J.} {\bfseries C74} no.~8, (2014) 3024},
\href{http://arxiv.org/abs/1404.5630}{{\ttfamily arXiv:1404.5630 [hep-ph]}}.
%%CITATION = ARXIV:1404.5630;%%.

\bibitem{GEANT3}
R.~Brun {\em et~al.}, ``{GEANT Detector Description and Simulation Tool},''
{\em CERN-W5013, CERN-W-5013, W5013, W-5013} (1994) .
%%CITATION = CERN-W5013;%%.

\bibitem{Lange:2001uf}
D.~Lange, ``{The EvtGen particle decay simulation package},''
\href{http://dx.doi.org/10.1016/S0168-9002(01)00089-4}{{\em Nucl. Instrum.
  Meth. A} {\bfseries 462} (2001) 152--155}.
%%CITATION = NUIMA,A462,152;%%.

\bibitem{Barberio:1993qi}
E.~Barberio and Z.~Was, ``{PHOTOS: A Universal Monte Carlo for QED radiative
  corrections. Version 2.0},''
\href{http://dx.doi.org/10.1016/0010-4655(94)90074-4}{{\em Comput. Phys.
  Commun.} {\bfseries 79} (1994) 291--308}.
%%CITATION = CPHCB,79,291;%%.

\bibitem{Acharya:2018uww}
{\bfseries ALICE} Collaboration, S.~Acharya {\em et~al.}, ``{Measurement of the
  inclusive J/ $\psi $ polarization at forward rapidity in pp collisions at
  $\mathbf {\sqrt{s} = 8}$ TeV},''
  \href{http://dx.doi.org/10.1140/epjc/s10052-018-6027-2}{{\em Eur. Phys. J.}
  {\bfseries C78} no.~7, (2018) 562},
\href{http://arxiv.org/abs/1805.04374}{{\ttfamily arXiv:1805.04374 [hep-ex]}}.
%%CITATION = ARXIV:1805.04374;%%.

\bibitem{Malaescu:2009dm}
B.~Malaescu, ``{An Iterative, dynamically stabilized method of data
  unfolding},''
\href{http://arxiv.org/abs/0907.3791}{{\ttfamily arXiv:0907.3791
  [physics.data-an]}}.
%%CITATION = ARXIV:0907.3791;%%.

\bibitem{Abelev:2012gx}
{\bfseries ALICE} Collaboration, B.~Abelev {\em et~al.}, ``{Measurement of
  prompt $J/\psi$ and beauty hadron production cross sections at mid-rapidity
  in pp collisions at $\sqrt{s} = 7$ TeV},''
  \href{http://dx.doi.org/10.1007/JHEP11(2012)065}{{\em JHEP} {\bfseries 11}
  (2012) 065},
\href{http://arxiv.org/abs/1205.5880}{{\ttfamily arXiv:1205.5880 [hep-ex]}}.
%%CITATION = ARXIV:1205.5880;%%.

\bibitem{Acharya:2019kyh}
{\bfseries ALICE} Collaboration, S.~Acharya {\em et~al.}, ``{Multiplicity
  dependence of (multi-)strange hadron production in proton-proton collisions
  at $\sqrt{s}$ = 13 TeV},''
  \href{http://dx.doi.org/10.1140/epjc/s10052-020-7673-8}{{\em Eur. Phys. J.}
  {\bfseries C80} no.~2, (2020) 167},
\href{http://arxiv.org/abs/1908.01861}{{\ttfamily arXiv:1908.01861 [nucl-ex]}}.
%%CITATION = ARXIV:1908.01861;%%.

\bibitem{Werner:2010aa}
K.~Werner, I.~Karpenko, T.~Pierog, M.~Bleicher, and K.~Mikhailov,
  ``{Event-by-Event Simulation of the Three-Dimensional Hydrodynamic Evolution
  from Flux Tube Initial Conditions in Ultrarelativistic Heavy Ion
  Collisions},'' \href{http://dx.doi.org/10.1103/PhysRevC.82.044904}{{\em Phys.
  Rev.} {\bfseries C82} (2010) 044904},
\href{http://arxiv.org/abs/1004.0805}{{\ttfamily arXiv:1004.0805 [nucl-th]}}.
%%CITATION = ARXIV:1004.0805;%%.

\bibitem{Drescher:2000ha}
H.~J. Drescher, M.~Hladik, S.~Ostapchenko, T.~Pierog, and K.~Werner, ``{Parton
  based Gribov-Regge theory},''
  \href{http://dx.doi.org/10.1016/S0370-1573(00)00122-8}{{\em Phys. Rept.}
  {\bfseries 350} (2001) 93--289},
\href{http://arxiv.org/abs/hep-ph/0007198}{{\ttfamily arXiv:hep-ph/0007198
  [hep-ph]}}.
%%CITATION = HEP-PH/0007198;%%.

\bibitem{Kopeliovich:2019phc}
B.~Z. Kopeliovich, H.~J. Pirner, I.~K. Potashnikova, K.~Reygers, and
  I.~Schmidt, ``{Heavy quarkonium in the saturated environment of
  high-multiplicity pp collisions},''
  \href{http://dx.doi.org/10.1103/PhysRevD.101.054023}{{\em Phys. Rev. D}
  {\bfseries 101} no.~5, (2020) 054023},
  \href{http://arxiv.org/abs/1910.09682}{{\ttfamily arXiv:1910.09682
  [hep-ph]}}.

\end{thebibliography}\endgroup
%\input {bibliography.tex}  

%%%%%%%%% appendix with author list
\newpage
\appendix
\section{The ALICE Collaboration} \label{app:collab}
% Collaboration: CERN-LHC-ALICE
% Generation Date is 2020-05-07

% How to use:
%%%%%%%%% appendix with author list
%\appendix
%\section{The ALICE Collaboration}
%\label{app:collab}
%\input{Alice_Authorslist_XXXX-Axx-XX.tex}
\begingroup
\small
\begin{flushleft}
S.~Acharya\Irefn{org141}\And 
D.~Adamov\'{a}\Irefn{org95}\And 
A.~Adler\Irefn{org74}\And 
J.~Adolfsson\Irefn{org81}\And 
M.M.~Aggarwal\Irefn{org100}\And 
G.~Aglieri Rinella\Irefn{org34}\And 
M.~Agnello\Irefn{org30}\And 
N.~Agrawal\Irefn{org10}\textsuperscript{,}\Irefn{org54}\And 
Z.~Ahammed\Irefn{org141}\And 
S.~Ahmad\Irefn{org16}
% ...
\And
\renewcommand\labelenumi{\textsuperscript{\theenumi}~}

\section*{Affiliation notes}
\renewcommand\theenumi{\roman{enumi}}
\begin{Authlist}
\item \Adef{org*}Deceased
\item \Adef{orgI}Italian National Agency for New Technologies, Energy and Sustainable Economic Development (ENEA), Bologna, Italy
\item \Adef{orgII}Dipartimento DET del Politecnico di Torino, Turin, Italy
\item \Adef{orgIII}M.V. Lomonosov Moscow State University, D.V. Skobeltsyn Institute of Nuclear, Physics, Moscow, Russia
\item \Adef{orgIV}Department of Applied Physics, Aligarh Muslim University, Aligarh, India
\item \Adef{orgV}Institute of Theoretical Physics, University of Wroclaw, Poland
\end{Authlist}

\section*{Collaboration Institutes}
\renewcommand\theenumi{\arabic{enumi}~}
\begin{Authlist}
\item \Idef{org1}A.I. Alikhanyan National Science Laboratory (Yerevan Physics Institute) Foundation, Yerevan, Armenia
\item \Idef{org2}Bogolyubov Institute for Theoretical Physics, National Academy of Sciences of Ukraine, Kiev, Ukraine
\item \Idef{org3}Bose Institute, Department of Physics  and Centre for Astroparticle Physics and Space Science (CAPSS), Kolkata, India
\item \Idef{org4}Budker Institute for Nuclear Physics, Novosibirsk, Russia
\item \Idef{org5}California Polytechnic State University, San Luis Obispo, California, United States
\item \Idef{org6}Central China Normal University, Wuhan, China
\item \Idef{org7}Centre de Calcul de l'IN2P3, Villeurbanne, Lyon, France
\item \Idef{org8}Centro de Aplicaciones Tecnol\'{o}gicas y Desarrollo Nuclear (CEADEN), Havana, Cuba
\item \Idef{org9}Centro de Investigaci\'{o}n y de Estudios Avanzados (CINVESTAV), Mexico City and M\'{e}rida, Mexico
\item \Idef{org10}Centro Fermi - Museo Storico della Fisica e Centro Studi e Ricerche ``Enrico Fermi', Rome, Italy
\item \Idef{org11}Chicago State University, Chicago, Illinois, United States
\item \Idef{org12}China Institute of Atomic Energy, Beijing, China
\item \Idef{org13}Comenius University Bratislava, Faculty of Mathematics, Physics and Informatics, Bratislava, Slovakia
\item \Idef{org14}COMSATS University Islamabad, Islamabad, Pakistan
\item \Idef{org15}Creighton University, Omaha, Nebraska, United States
\item \Idef{org16}Department of Physics, Aligarh Muslim University, Aligarh, India
\item \Idef{org17}Department of Physics, Pusan National University, Pusan, Republic of Korea
\item \Idef{org18}Department of Physics, Sejong University, Seoul, Republic of Korea
\item \Idef{org19}Department of Physics, University of California, Berkeley, California, United States
\item \Idef{org20}Department of Physics, University of Oslo, Oslo, Norway
\item \Idef{org21}Department of Physics and Technology, University of Bergen, Bergen, Norway
\item \Idef{org22}Dipartimento di Fisica dell'Universit\`{a} 'La Sapienza' and Sezione INFN, Rome, Italy
\item \Idef{org23}Dipartimento di Fisica dell'Universit\`{a} and Sezione INFN, Cagliari, Italy
\item \Idef{org24}Dipartimento di Fisica dell'Universit\`{a} and Sezione INFN, Trieste, Italy
\item \Idef{org25}Dipartimento di Fisica dell'Universit\`{a} and Sezione INFN, Turin, Italy
\item \Idef{org26}Dipartimento di Fisica e Astronomia dell'Universit\`{a} and Sezione INFN, Bologna, Italy
\item \Idef{org27}Dipartimento di Fisica e Astronomia dell'Universit\`{a} and Sezione INFN, Catania, Italy
\item \Idef{org28}Dipartimento di Fisica e Astronomia dell'Universit\`{a} and Sezione INFN, Padova, Italy
\item \Idef{org29}Dipartimento di Fisica `E.R.~Caianiello' dell'Universit\`{a} and Gruppo Collegato INFN, Salerno, Italy
\item \Idef{org30}Dipartimento DISAT del Politecnico and Sezione INFN, Turin, Italy
\item \Idef{org31}Dipartimento di Scienze e Innovazione Tecnologica dell'Universit\`{a} del Piemonte Orientale and INFN Sezione di Torino, Alessandria, Italy
\item \Idef{org32}Dipartimento di Scienze MIFT, Universit\`{a} di Messina, Messina, Italy
\item \Idef{org33}Dipartimento Interateneo di Fisica `M.~Merlin' and Sezione INFN, Bari, Italy
\item \Idef{org34}European Organization for Nuclear Research (CERN), Geneva, Switzerland
\item \Idef{org35}Faculty of Electrical Engineering, Mechanical Engineering and Naval Architecture, University of Split, Split, Croatia
\item \Idef{org36}Faculty of Engineering and Science, Western Norway University of Applied Sciences, Bergen, Norway
\item \Idef{org37}Faculty of Nuclear Sciences and Physical Engineering, Czech Technical University in Prague, Prague, Czech Republic
\item \Idef{org38}Faculty of Science, P.J.~\v{S}af\'{a}rik University, Ko\v{s}ice, Slovakia
\item \Idef{org39}Frankfurt Institute for Advanced Studies, Johann Wolfgang Goethe-Universit\"{a}t Frankfurt, Frankfurt, Germany
\item \Idef{org40}Fudan University, Shanghai, China
\item \Idef{org41}Gangneung-Wonju National University, Gangneung, Republic of Korea
\item \Idef{org42}Gauhati University, Department of Physics, Guwahati, India
\item \Idef{org43}Helmholtz-Institut f\"{u}r Strahlen- und Kernphysik, Rheinische Friedrich-Wilhelms-Universit\"{a}t Bonn, Bonn, Germany
\item \Idef{org44}Helsinki Institute of Physics (HIP), Helsinki, Finland
\item \Idef{org45}High Energy Physics Group,  Universidad Aut\'{o}noma de Puebla, Puebla, Mexico
\item \Idef{org46}Hiroshima University, Hiroshima, Japan
\item \Idef{org47}Hochschule Worms, Zentrum  f\"{u}r Technologietransfer und Telekommunikation (ZTT), Worms, Germany
\item \Idef{org48}Horia Hulubei National Institute of Physics and Nuclear Engineering, Bucharest, Romania
\item \Idef{org49}Indian Institute of Technology Bombay (IIT), Mumbai, India
\item \Idef{org50}Indian Institute of Technology Indore, Indore, India
\item \Idef{org51}Indonesian Institute of Sciences, Jakarta, Indonesia
\item \Idef{org52}INFN, Laboratori Nazionali di Frascati, Frascati, Italy
\item \Idef{org53}INFN, Sezione di Bari, Bari, Italy
\item \Idef{org54}INFN, Sezione di Bologna, Bologna, Italy
\item \Idef{org55}INFN, Sezione di Cagliari, Cagliari, Italy
\item \Idef{org56}INFN, Sezione di Catania, Catania, Italy
\item \Idef{org57}INFN, Sezione di Padova, Padova, Italy
\item \Idef{org58}INFN, Sezione di Roma, Rome, Italy
\item \Idef{org59}INFN, Sezione di Torino, Turin, Italy
\item \Idef{org60}INFN, Sezione di Trieste, Trieste, Italy
\item \Idef{org61}Inha University, Incheon, Republic of Korea
\item \Idef{org62}Institute for Nuclear Research, Academy of Sciences, Moscow, Russia
\item \Idef{org63}Institute for Subatomic Physics, Utrecht University/Nikhef, Utrecht, Netherlands
\item \Idef{org64}Institute of Experimental Physics, Slovak Academy of Sciences, Ko\v{s}ice, Slovakia
\item \Idef{org65}Institute of Physics, Homi Bhabha National Institute, Bhubaneswar, India
\item \Idef{org66}Institute of Physics of the Czech Academy of Sciences, Prague, Czech Republic
\item \Idef{org67}Institute of Space Science (ISS), Bucharest, Romania
\item \Idef{org68}Institut f\"{u}r Kernphysik, Johann Wolfgang Goethe-Universit\"{a}t Frankfurt, Frankfurt, Germany
\item \Idef{org69}Instituto de Ciencias Nucleares, Universidad Nacional Aut\'{o}noma de M\'{e}xico, Mexico City, Mexico
\item \Idef{org70}Instituto de F\'{i}sica, Universidade Federal do Rio Grande do Sul (UFRGS), Porto Alegre, Brazil
\item \Idef{org71}Instituto de F\'{\i}sica, Universidad Nacional Aut\'{o}noma de M\'{e}xico, Mexico City, Mexico
\item \Idef{org72}iThemba LABS, National Research Foundation, Somerset West, South Africa
\item \Idef{org73}Jeonbuk National University, Jeonju, Republic of Korea
\item \Idef{org74}Johann-Wolfgang-Goethe Universit\"{a}t Frankfurt Institut f\"{u}r Informatik, Fachbereich Informatik und Mathematik, Frankfurt, Germany
\item \Idef{org75}Joint Institute for Nuclear Research (JINR), Dubna, Russia
\item \Idef{org76}Korea Institute of Science and Technology Information, Daejeon, Republic of Korea
\item \Idef{org77}KTO Karatay University, Konya, Turkey
\item \Idef{org78}Laboratoire de Physique des 2 Infinis, Ir\`{e}ne Joliot-Curie, Orsay, France
\item \Idef{org79}Laboratoire de Physique Subatomique et de Cosmologie, Universit\'{e} Grenoble-Alpes, CNRS-IN2P3, Grenoble, France
\item \Idef{org80}Lawrence Berkeley National Laboratory, Berkeley, California, United States
\item \Idef{org81}Lund University Department of Physics, Division of Particle Physics, Lund, Sweden
\item \Idef{org82}Nagasaki Institute of Applied Science, Nagasaki, Japan
\item \Idef{org83}Nara Women{'}s University (NWU), Nara, Japan
\item \Idef{org84}National and Kapodistrian University of Athens, School of Science, Department of Physics , Athens, Greece
\item \Idef{org85}National Centre for Nuclear Research, Warsaw, Poland
\item \Idef{org86}National Institute of Science Education and Research, Homi Bhabha National Institute, Jatni, India
\item \Idef{org87}National Nuclear Research Center, Baku, Azerbaijan
\item \Idef{org88}National Research Centre Kurchatov Institute, Moscow, Russia
\item \Idef{org89}Niels Bohr Institute, University of Copenhagen, Copenhagen, Denmark
\item \Idef{org90}Nikhef, National institute for subatomic physics, Amsterdam, Netherlands
\item \Idef{org91}NRC Kurchatov Institute IHEP, Protvino, Russia
\item \Idef{org92}NRC \guillemotleft Kurchatov\guillemotright~Institute - ITEP, Moscow, Russia
\item \Idef{org93}NRNU Moscow Engineering Physics Institute, Moscow, Russia
\item \Idef{org94}Nuclear Physics Group, STFC Daresbury Laboratory, Daresbury, United Kingdom
\item \Idef{org95}Nuclear Physics Institute of the Czech Academy of Sciences, \v{R}e\v{z} u Prahy, Czech Republic
\item \Idef{org96}Oak Ridge National Laboratory, Oak Ridge, Tennessee, United States
\item \Idef{org97}Ohio State University, Columbus, Ohio, United States
\item \Idef{org98}Petersburg Nuclear Physics Institute, Gatchina, Russia
\item \Idef{org99}Physics department, Faculty of science, University of Zagreb, Zagreb, Croatia
\item \Idef{org100}Physics Department, Panjab University, Chandigarh, India
\item \Idef{org101}Physics Department, University of Jammu, Jammu, India
\item \Idef{org102}Physics Department, University of Rajasthan, Jaipur, India
\item \Idef{org103}Physikalisches Institut, Eberhard-Karls-Universit\"{a}t T\"{u}bingen, T\"{u}bingen, Germany
\item \Idef{org104}Physikalisches Institut, Ruprecht-Karls-Universit\"{a}t Heidelberg, Heidelberg, Germany
\item \Idef{org105}Physik Department, Technische Universit\"{a}t M\"{u}nchen, Munich, Germany
\item \Idef{org106}Politecnico di Bari, Bari, Italy
\item \Idef{org107}Research Division and ExtreMe Matter Institute EMMI, GSI Helmholtzzentrum f\"ur Schwerionenforschung GmbH, Darmstadt, Germany
\item \Idef{org108}Rudjer Bo\v{s}kovi\'{c} Institute, Zagreb, Croatia
\item \Idef{org109}Russian Federal Nuclear Center (VNIIEF), Sarov, Russia
\item \Idef{org110}Saha Institute of Nuclear Physics, Homi Bhabha National Institute, Kolkata, India
\item \Idef{org111}School of Physics and Astronomy, University of Birmingham, Birmingham, United Kingdom
\item \Idef{org112}Secci\'{o}n F\'{\i}sica, Departamento de Ciencias, Pontificia Universidad Cat\'{o}lica del Per\'{u}, Lima, Peru
\item \Idef{org113}St. Petersburg State University, St. Petersburg, Russia
\item \Idef{org114}Stefan Meyer Institut f\"{u}r Subatomare Physik (SMI), Vienna, Austria
\item \Idef{org115}SUBATECH, IMT Atlantique, Universit\'{e} de Nantes, CNRS-IN2P3, Nantes, France
\item \Idef{org116}Suranaree University of Technology, Nakhon Ratchasima, Thailand
\item \Idef{org117}Technical University of Ko\v{s}ice, Ko\v{s}ice, Slovakia
\item \Idef{org118}The Henryk Niewodniczanski Institute of Nuclear Physics, Polish Academy of Sciences, Cracow, Poland
\item \Idef{org119}The University of Texas at Austin, Austin, Texas, United States
\item \Idef{org120}Universidad Aut\'{o}noma de Sinaloa, Culiac\'{a}n, Mexico
\item \Idef{org121}Universidade de S\~{a}o Paulo (USP), S\~{a}o Paulo, Brazil
\item \Idef{org122}Universidade Estadual de Campinas (UNICAMP), Campinas, Brazil
\item \Idef{org123}Universidade Federal do ABC, Santo Andre, Brazil
\item \Idef{org124}University of Cape Town, Cape Town, South Africa
\item \Idef{org125}University of Houston, Houston, Texas, United States
\item \Idef{org126}University of Jyv\"{a}skyl\"{a}, Jyv\"{a}skyl\"{a}, Finland
\item \Idef{org127}University of Liverpool, Liverpool, United Kingdom
\item \Idef{org128}University of Science and Technology of China, Hefei, China
\item \Idef{org129}University of South-Eastern Norway, Tonsberg, Norway
\item \Idef{org130}University of Tennessee, Knoxville, Tennessee, United States
\item \Idef{org131}University of the Witwatersrand, Johannesburg, South Africa
\item \Idef{org132}University of Tokyo, Tokyo, Japan
\item \Idef{org133}University of Tsukuba, Tsukuba, Japan
\item \Idef{org134}Universit\'{e} Clermont Auvergne, CNRS/IN2P3, LPC, Clermont-Ferrand, France
\item \Idef{org135}Universit\'{e} de Lyon, Universit\'{e} Lyon 1, CNRS/IN2P3, IPN-Lyon, Villeurbanne, Lyon, France
\item \Idef{org136}Universit\'{e} de Strasbourg, CNRS, IPHC UMR 7178, F-67000 Strasbourg, France, Strasbourg, France
\item \Idef{org137}Universit\'{e} Paris-Saclay Centre d'Etudes de Saclay (CEA), IRFU, D\'{e}partment de Physique Nucl\'{e}aire (DPhN), Saclay, France
\item \Idef{org138}Universit\`{a} degli Studi di Foggia, Foggia, Italy
\item \Idef{org139}Universit\`{a} degli Studi di Pavia, Pavia, Italy
\item \Idef{org140}Universit\`{a} di Brescia, Brescia, Italy
\item \Idef{org141}Variable Energy Cyclotron Centre, Homi Bhabha National Institute, Kolkata, India
\item \Idef{org142}Warsaw University of Technology, Warsaw, Poland
\item \Idef{org143}Wayne State University, Detroit, Michigan, United States
\item \Idef{org144}Westf\"{a}lische Wilhelms-Universit\"{a}t M\"{u}nster, Institut f\"{u}r Kernphysik, M\"{u}nster, Germany
\item \Idef{org145}Wigner Research Centre for Physics, Budapest, Hungary
\item \Idef{org146}Yale University, New Haven, Connecticut, United States
\item \Idef{org147}Yonsei University, Seoul, Republic of Korea
\end{Authlist}
\endgroup
  %%%%%%% done by webmaster team

\end{document}